\documentclass[preprint,12pt,authoryear]{elsarticle}

\usepackage{amssymb}
\usepackage{amsthm}
\usepackage{amsmath} 
\usepackage{subfig}
\usepackage{multirow}

\usepackage{xurl} 
\usepackage{lscape} 
\usepackage[table,xcdraw]{xcolor} 

\journal{Journal of Banking \& Finance}

\begin{document}

\begin{frontmatter}



\title{Using CPI in Loss Given Default Forecasting Models for Commercial Real Estate Portfolio \tnoteref{t1,t2}}


\author[1]{Ying Wu\fnref{fn1}}
\ead{ying.wu@jpmchase.com}

\author[2]{Garvit Arora\fnref{fn1}}
\ead{arora.garvit@jpmchase.com}

\author[1]{Xuan Mei\corref{cor1}\fnref{fn1}}
\ead{xuan.mei@chase.com}

\cortext[cor1]{Corresponding author}

\fntext[fn1]{Ying, Garvit and Xuan are all working at the Wholesale Credit QR group in JPMorgan Chase \& Co..}
\affiliation[1]{organization={JPMorgan Chase \& Co.},
	addressline={545 Washington Blvd.}, 
	city={Jersey City},
	postcode={07310}, 
	state={NJ},
	country={USA}}

\affiliation[2]{organization={JPMorgan Chase \& Co.},
            addressline={Level 3 and 4 J.P. Mogan Tower, Off Cst Road Kalina Santacruz East}, 
            city={Mumbai},
            postcode={400098}, 
            country={India}}

\begin{abstract}
Forecasting the loss given default (LGD) for defaulted Commercial Real Estate (CRE) loans poses a significant challenge due to the extended resolution and workout time associated with such defaults, particularly in CCAR and CECL framework where the utilization of post-default information, including macroeconomic variables (MEVs) such as unemployment (UER) and various rates, is restricted. The current environment of persistent inflation and resultant elevated rates further compounds the uncertainty surrounding predictive LGD models. In this paper, we leverage both internal and public data sources, including observations from the COVID-19 period, to present a list of evidence indicating that the growth rates of the Consumer Price Index (CPI)\footnote{The official name, by U.S. Bureau of Labor Statistics, is Consumer Price Index for All Urban Consumers (CPI-U), U.S. City Average All items seasonally-adjusted index}, such as Year-over-Year (YoY) growth and logarithmic growth, are good leading indicators  for various CRE related rates and indices. These include the Federal Funds Effective Rate and CRE market sales price indices in key locations such as Los Angeles, New York, and nationwide, encompassing both apartment and office segments. Furthermore, with CRE LGD data we demonstrate how incorporating CPI at the time of default can improve the accuracy of predicting CRE workout LGD. This is particularly helpful in addressing the common issue of early downturn underestimation encountered in CRE LGD models. 
\end{abstract}




\begin{keyword}



Commercial real estate; CPI; LGD; early downturn underestimation; COVID-19
\end{keyword}

\end{frontmatter}


\section{Introduction}
\label{sec:intro}
\subsection{CRE CTL portfolio in JPMC}
The Commercial Term Lending (CTL) business within JPMorgan Chase (JPMC) engages in long-term, permanent mortgage financing on stabilized income producing commercial real estate (CRE). The typical loan is made to "mom and pop" borrowers who hold investment property for an extended period of time as a secondary source of income and has an average loan balance at origination of roughly \$2.4 million based on 2020 origination. The majority (90\%) of loans are for multifamily lending (MFL) property type; however, it has industrial (e.g., warehouse), office, retail, and other commercial mortgage lending (CML) property types. CTL is currently the nation's biggest multifamily lender. 

As a large bank holding, JPMC is required to participate in the U.S. Federal Reserve Board System (FRB) comprehensive capital analysis and review (CCAR) exercises as well as compute reserves for US GAAP under the current expected credit loss (CECL) standard. Please see  \cite{CECL2016} and \cite{CCAR2020} for more details.

It is Wholesale Credit QR team's (authors') responsibility to cover the CTL portfolio and develop a suite of loss forecast models  compliant to CCAR and CECL framework and requirements, including but not limited to the probability of default (PD) model and loss given default (LGD) model.

\subsection{Challenges in LGD modeling}
Compared to the PD, we found the LGD model more difficult to build because of two regularities. 

First, historical data shows that when a CTL borrower went default, after the usually long and costly foreclosure process, JPMC still had a good chance to see a positive net proceeds. Weather or not JPMC has the right to keep these proceeds depends on the specific contract and covenant binding, which means that sometimes JPMC could see a negative loss on defaulted CRE loans. However, by CECL and CCAR rules the LGD must be non-negative. In more accurate math language, what we are dealing with is not the raw LGD but a left censored version of it, i.e., $\text{LGD}^+ = \text{LGD}\cdot I(\text{LGD}>=0)$, where $I(\cdot)$ is the indicator function which takes value 1 if the condition in the parenthesis is met and 0 elsewhere.  This left censoring makes LGD a bi-modal distribution with non-zero mass at 0. Popular solutions include Tobit I (\cite{tobin1958estimation}) or Tobit II model (aka Heckit, \cite{heckman1979sample}), fractional response model (aka FRM, \cite{papke1996econometric}), inflated beta regression (\cite{ospina2010inflated}) censored gamma regression (\cite{sigrist2011using}). This is not our topic in this paper. 

\begin{figure}[h!]
         \caption{FRB Commercial Real Estate Prices, YoY \% Change }\label{fig:CREPI_YoY}
	\centering
	\includegraphics[scale=0.5]{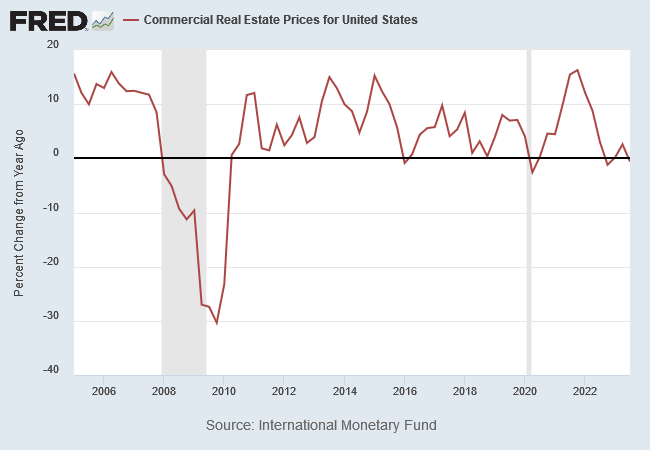}
\end{figure}

Second, it is well accepted that the inclusion of macroeconomic variables  (MEVs) such as UER (unemployment rate), GDP, HPI, etc., generally improves the LGD predictions (see e.g., \cite{bellotti2012loss}). However, for a default occurred at time $t$, when projecting the corresponding loss, CECL and CCAR framework only allow us use information up to $t$, and thus MEVs quantitatively depicting the economy after default are not allowed in LGD models.  For CRE defaults, the foreclosure or workout process can normally take 12 months or longer. Thus the value of collateral properties at final sale can deviate from what was appraised at or before default. From Figure \ref{fig:CREPI_YoY} (the U.S. Commercial Real Estate Price Index\footnote{This is known as CREPI, derived from CoStar Commercial Repeat-Sale Indices (CCRSI), aiming to capture the movement of CRE market actual transaction price. For details, please see \cite{CoStar_CCRSI}} YoY change, published by FRB), we see that for CRE loans defaulted at 2008, when the underlying properties finished the 12 months foreclosure process and came to the market for sale in 2009, the CRE market sales price already sunk by 10\% $\sim$ 30\%. This can cause excessive losses that can hardly be captured by most MEVs at default time, and causes an "early downturn LGD underestimation" issue in CRE LGD modeling. 

After exploration, we find CPI (Consumer Price Index for All Urban Consumers (CPI-U), U.S. City Average All items seasonally-adjusted index, see Section \ref{{sec:cpi_intro}}), as a good leading indicator of future rates and CRE market sales price, can be enlisted to address this "early downturn LGD underestimation" issue, at least true for the CTL portfolio we are covering. We notice that in all major LGD papers we have reviewed (see Section \ref{sec:literature}), few author had explored this topic nor did the y ever put CPI in their candidate variable pool. This could create some vulnerability in LGD models given nowadays persistent CPI and the looming worry about a stagflation. We want to share our findings in tandem with all evidences with researchers who are interested in this topic.

The rest of paper is structured as follows: Section 2 summarizes existing researches on LGD models we found relevant to our topic, particularly those using rates or CPI in their LGD models.  In Section 3 we present our findings about the relation between CPI and spot rates, future rates, unemployment and CRE market sales price, based on both public and JPMC internal data, with economic and business justifications. Section 4 lists statistical and machine learning evidences that prove the usefulness of adding CPI in our CRE CTL LGD model, including various cross validation and multivariate adaptive regression splines (MARS) used to prevent spurious regression and detect nonlinear patterns of CPI.  Section 5 concludes.


\section{Literature Review}
\label{sec:literature}

To our knowledge, using MEVs in LGD models started after GFC when FRB proposed the CCAR framework where future economy is depicted by a list of MEVs, such as UER, GDP, HPI and various interest rates including the Fed Fund Rate. \cite{bellotti2012loss} In Table \ref{table:literature}, we summarize all public LGD papers we could find using or discussing using MEVs, covering a wide range of portfolios, including residential mortgage, bonds, credit cards, bank loans and unsecured consumer loans from Lending Club (see Data column). The column Time and MEVs (sign) list respectively the window of time their data covers, MEVs they considered and the corresponding signs in their models. Note that, some authors studied recovery rate instead of LGD, and given $\text{LGD} = 1 - \text{recovery rate}$, their MEVs' signs should be flipped when talking about LGD.

\begin{table}[ht]
\caption{LGD and MEVs Literature Review Summary}\label{table:literature}
\scalebox{0.52}{
\begin{tabular}{|l|l|l|l|l|l|}
\hline
                                            & \textbf{Data}                                                                                                   & \textbf{Method}                                                                                                                        & \textbf{Country}                                                       & \textbf{Time}               & \textbf{MEVs (sign)}                                    \\ \hline
\cite{qi2009loss}                         & \begin{tabular}[c]{@{}l@{}}Residential mortgages\\ (workout \textbf{LGD})\end{tabular}                                   & Regression                                                                                                                             & US                                                                     & 1990-2003                   & Stress dummy (+)                                        \\ \hline
                                            &                                                                                                                 &                                                                                                                                        &                                                                        &                             & {\color[HTML]{3531FF} Current LTV (+)}                  \\ \cline{6-6} 
                                            &                                                                                                                 &                                                                                                                                        &                                                                        &                             & Industry distance to default (-)                        \\ \cline{6-6} 
                                            &                                                                                                                 &                                                                                                                                        &                                                                        &                             & Industry default rate (+)                               \\ \cline{6-6} 
                                            &                                                                                                                 &                                                                                                                                        &                                                                        &                             & Market return (-)                                       \\ \cline{6-6} 
\multirow{-5}{*}{\cite{qi2011comparison}}        & \multirow{-5}{*}{\begin{tabular}[c]{@{}l@{}}Bonds \\ (market-based and \\ workout \textbf{LGD})\end{tabular}}            & \multirow{-5}{*}{\begin{tabular}[c]{@{}l@{}}6 methods, \\ including Tobit, \\ Inverse Gaussian, \\ inflated beta, \\ FRM\end{tabular}} & \multirow{-5}{*}{US}                                                   & \multirow{-5}{*}{1985-2008} & {\color[HTML]{3531FF} \textbf{Interest rate (+)}}       \\ \hline
                                            &                                                                                                                 &                                                                                                                                        &                                                                        &                             & {\color[HTML]{3531FF} \textbf{Interest rate (-)}}       \\ \cline{6-6} 
\multirow{-2}{*}{\cite{bellotti2012loss}} & \multirow{-2}{*}{\begin{tabular}[c]{@{}l@{}}Personal credit cards \\ (workout \textbf{recovery rate})\end{tabular}}      & \multirow{-2}{*}{\begin{tabular}[c]{@{}l@{}}Regression tree, \\ Tobit I\end{tabular}}                                                  & \multirow{-2}{*}{UK}                                                   & \multirow{-2}{*}{1999-2005} & {\color[HTML]{3531FF} UER (-)}                 \\ \hline
                                            &                                                                                                                 &                                                                                                                                        &                                                                        &                             & Market default rate (-)                                 \\ \cline{6-6} 
                                            &                                                                                                                 &                                                                                                                                        &                                                                        &                             & Industry default rate (-)                               \\ \cline{6-6} 
\multirow{-3}{*}{\cite{jankowitsch2014determinants}} & \multirow{-3}{*}{\begin{tabular}[c]{@{}l@{}}Bonds \\ (market-based \textbf{recovery rate})\end{tabular}}                 & \multirow{-3}{*}{Regression}                                                                                                           & \multirow{-3}{*}{US}                                                   & \multirow{-3}{*}{2002-2010} & {\color[HTML]{FE0000} \textbf{Federal funds rate (+)}}  \\ \hline
                                            &                                                                                                                 &                                                                                                                                        &                                                                        &                             & {\color[HTML]{FE0000} UER (+)}                  \\ \cline{6-6} 
                                            &                                                                                                                 &                                                                                                                                        &                                                                        &                             & {\color[HTML]{3531FF} \textbf{CPI (-)}}                 \\ \cline{6-6} 
\multirow{-3}{*}{\cite{yao2017enhancing}}         & \multirow{-3}{*}{\begin{tabular}[c]{@{}l@{}}Credit cards \\ (workout \textbf{recovery rate})\end{tabular}}                    & \multirow{-3}{*}{\begin{tabular}[c]{@{}l@{}}Two stage model, \\ with Support Vector \\ Machine\end{tabular}}                           & \multirow{-3}{*}{UK}                                                   & \multirow{-3}{*}{2009-2010} & {\color[HTML]{FE0000} HPI (-)}                          \\ \hline
                                            &                                                                                                                 &                                                                                                                                        &                                                                        &                             & {\color[HTML]{3531FF} GDP (-)}                          \\ \cline{6-6} 
                                            &                                                                                                                 &                                                                                                                                        &                                                                        &                             & {\color[HTML]{3531FF} EI (-)}                           \\ \cline{6-6} 
                                            &                                                                                                                 &                                                                                                                                        &                                                                        &                             & {\color[HTML]{3531FF} \textbf{VIX (+)}}                 \\ \cline{6-6} 
\multirow{-4}{*}{\cite{betz2018systematic}}      & \multirow{-4}{*}{\begin{tabular}[c]{@{}l@{}}Bank loans \\ from Global Credit Data\\ (workout \textbf{LGD})\end{tabular}} & \multirow{-4}{*}{\begin{tabular}[c]{@{}l@{}}hierarchical model \\ combining Finite Mixture\\ Model\end{tabular}}                       & \multirow{-4}{*}{\begin{tabular}[c]{@{}l@{}}US\\ UK\\ EU\end{tabular}} & \multirow{-4}{*}{2006-2012} & {\color[HTML]{3531FF} HPI (-)}                          \\ \hline
                                            &                                                                                                                 &                                                                                                                                        &                                                                        &                             & {\color[HTML]{3531FF} \textbf{Prime interest rate (+)}} \\ \cline{6-6} 
                                            &                                                                                                                 &                                                                                                                                        &                                                                        &                             & Producer price index (-)                                \\ \cline{6-6} 
                                            &                                                                                                                 &                                                                                                                                        &                                                                        &                             & {\color[HTML]{3531FF} UER (+)}                 \\ \cline{6-6} 
\multirow{-4}{*}{\cite{li2023predicting}}                  & \multirow{-4}{*}{\begin{tabular}[c]{@{}l@{}}Unsecured consumer loans \\ from Lending Club\\ (workout \textbf{LGD})\end{tabular}}         & \multirow{-4}{*}{\begin{tabular}[c]{@{}l@{}}time varying coefficients,\\ with Cox proportional \\ hazard model\end{tabular}}           & \multirow{-4}{*}{US}                                                   & \multirow{-4}{*}{2016-2019} & Average real wage index (-)                             \\ \hline
\end{tabular}
}
\end{table}

\subsection{LGD and Rates}
\label{sec:lgd_rates}
For interest rates, \cite{qi2011comparison} (US bonds workout LGD, 1985-2008), \cite{bellotti2012loss} (UL personal credit card LGD, 1999-2005) and \cite{li2023predicting} (Lending Club consumer loans, 2016-2019) all reported positive signs of interest rates in their LGD models, which implies that a higher interest rate causes higher LGD. However, none of them has data coverage over the whole 2008-2010 GFC period. And the one who has that, i.e., \cite{jankowitsch2014determinants},  reported positive Federal funds rate towards bonds recovery rate in their US bonds market-based recovery rate model, which implies a decreasing Federal funds rate will increase bonds LGD, with all other inputs held the same. Their explanation is that, FRB only lowers rates when the economy is under stress and increases it when the economy is performing well, and thus a high interest rate means good economy, which leads to low bonds LGD. \cite{betz2018systematic} (banks loans workout LGD in US, UK, EU, 2006 - 2012) did not consider interest rates, instead they found EI (the quarterly average of YoY log returns of major stock indices\footnote{They use the S\&P500 for US and FTSE for UK}) negatively correlated to LGD, while VIX, the fear index positively correlated to LGD.    

This in certain way explains what we encounter in our CRE CTL LGD model data spanning from 2004 to 2019. We tried various treasury rates and spreads, and their transformations such as 1/2/3/4Q simple differences. However, in Tobit I regression model, they are either insignificant or having negative signs. Given the nowadays high interest rate environment, and the fear of a stagflation crisis where interest rates will stay high, we are not feeling comfortable admitting a LGD model with negative coefficients on rates. 

\subsection{LGD and CPI}
There is one particular literature, \cite{yao2017enhancing}, that specifically uses CPI as a risk driver for workout recovery rates with negative sign, though it was in UK credit card space. The negative sign between recovery rate and CPI indicates a higher LGD as CPI increases, which is in line with our positive sign for CPI. This observation is interpreted as high inflation will quickly drain consumers saving and make collection more difficult. Given the backbone of CTL portfolio are small investors, it is reasonable to believe that high inflation will also erode their purchasing power for our foreclosed CRE properties for sale. We want to point out that in their paper, UER is found to be positively correlated to recovery rate while HPI is negatively correlation to recovery rate. The authors are also puzzled by the unintuitive sign of UER which is contradicting to \cite{bellotti2012loss}, and they think it could be due to the short span of their data. We have a guess that the way they use HPI and CPI, i.e., using the non-stationary levels instead of their stationary transformation perhaps also contributes to the spurious signs.

This is the only literature we found that directly uses CPI. However, this is due to this literature is the only one which explicitly considers CPI in its variable pool. Inflation had been very stable for decades and was probability not in the scope of many modelers and researches, so literature in this area usually do not even consider CPI in variable selection pool. It is hard to tell whether CPI would be used if they added it into candidate pool, since we found CPI has close relation to various interest rates, as well as the CRE market sale price indices.

\section{CPI, Rates and CRE Market Sales Price Indices} 
\label{sec:cpi_rate}

\subsection{Various CPI transformations}
\label{sec:cpi_intro}
The official name of the CPI we have been mentioning throughout this paper is, by U.S. Bureau of Labor Statistics (BLS), Consumer Price Index for All Urban Consumers (CPI-U), U.S. City Average All items seasonally-adjusted index. As a weighted average of prices for a basket of goods and services representative of aggregated U.S. consumer spending, this index is published monthly by BLS, using prices (taxes included) collected in 75 urban areas from about 6,000 housing units  and 22,000 retails establishment, covering 93\% of the U.S. population. It covers major groups of consumer expenditures such as food and beverages, housing, apparel, transportation, medical care, recreation, education and communications, and other goods and services, taxes included. Though derived from price changes and aiming to measure the change in prices paid by consumers for goods and services, this index itself is a level, based at 1982-1984 as 1, and around 3 today. Its historical data begin as early as 1913, but in JPMC we generally use this index after 1950. There are several variations of CPI, such as CPI-W, Core-CPI, C-CPI-U, etc. For a comprehensive introduction about CPI and its sampling methodology, one can refer to the CPI Technical Note \footnote{https://www.bls.gov/cpi/technical-notes/}. BLS highlights the importance of CPI "\textit{As an economic indicator. As the most widely used measure of inflation, the CPI is an indicator of the effectiveness of government policy. In addition, business executives, labor leaders and other private citizens use the index as a guide in making economic decisions}".

Given CPI is a level, we generally use its RDIFF$k$M and LDIFF$k$M, i.e., the $k$ months ratio difference and log difference, where $k$ can be 3, 6 and 12. We seldom use $k = 9$ because it is less common compared to the other three. $k=12$ gives the very common YoY and YoY log growth.  Also, for a chosen $k$, RDIFF$k$M and LDIFF$k$m are nearly identical, because of the simple relation between ratio difference $r$ and log difference $\log(1+r)$, i.e., 
$$\log(1+r) = r + O(r^2)$$
Table \ref{table:CPI_Corr} lists the correlation matrix of above mentioned transformations of CPI, based on data from 1950s to 2022. Given the extremely high correlation between them, in GLM or any model family where (row) vector inputs $X$ impact target variable $Y$ through a linear predictor $X\beta$, we think these transformations will give similar explanation power. For simplicity, in rest of this section we use the popular CPI YoY, i.e., CPI RDIFF12M throughout the analyses. 

\begin{table}[h]
\center
\caption{Correlation between CPI Transformations}\label{table:CPI_Corr}
\scalebox{0.45}{
\begin{tabular}{|l|c|c|c|c|c|c|}
\hline
                         & \multicolumn{1}{l|}{\textbf{CPI\_M.RDIFF12M}} & \multicolumn{1}{l|}{\textbf{CPI\_M.LDIFF12M}} & \multicolumn{1}{l|}{\textbf{CPI\_M.RDIFF6M}} & \multicolumn{1}{l|}{\textbf{CPI\_M.LDIFF6M}} & \multicolumn{1}{l|}{\textbf{CPI\_M.RDIFF3M}} & \multicolumn{1}{l|}{\textbf{CPI\_M.LDIFF3M}} \\ \hline
\textbf{CPI\_M.RDIFF12M} & 1                                             & 0.9998                                        & 0.9243                                       & 0.9228                                       & 0.8222                                       & 0.8205                                       \\ \hline
\textbf{CPI\_M.LDIFF12M} & 0.9998                                        & 1                                             & 0.9240                                       & 0.9226                                       & 0.8220                                       & 0.8204                                       \\ \hline
\textbf{CPI\_M.RDIFF6M}  & 0.9243                                        & 0.9240                                        & 1                                            & 0.9999                                       & 0.8967                                       & 0.8957                                       \\ \hline
\textbf{CPI\_M.LDIFF6M}  & 0.9228                                        & 0.9226                                        & 0.9999                                       & 1                                            & 0.8966                                       & 0.8957                                       \\ \hline
\textbf{CPI\_M.RDIFF3M}  & 0.8222                                        & 0.8220                                        & 0.8967                                       & 0.8966                                       & 1                                            & 0.99996                                      \\ \hline
\textbf{CPI\_M.LDIFF3M}  & 0.8205                                        & 0.8204                                        & 0.8957                                       & 0.8957                                       & 0.99996                                      & 1                                            \\ \hline
\end{tabular}
}
\end{table}

\subsection{CPI and spot rates}
We know fighting inflation is one of FRB's two mandates. It is intuitive that high CPI is a preamble of high rates: When CPI raises sharply (i.e., big CPI YoY), FRB is more likely to increase rates and hold them at high level for a period of time, making it more expensive for people to borrow funds to buy properties, which can potentially lower properties' value.  
\begin{figure}[ht]
	\caption {CPI YoY vs. Fed Fund Rate}\label{fig:CPI_fundrate}
	\centering
	\includegraphics[scale=0.295]{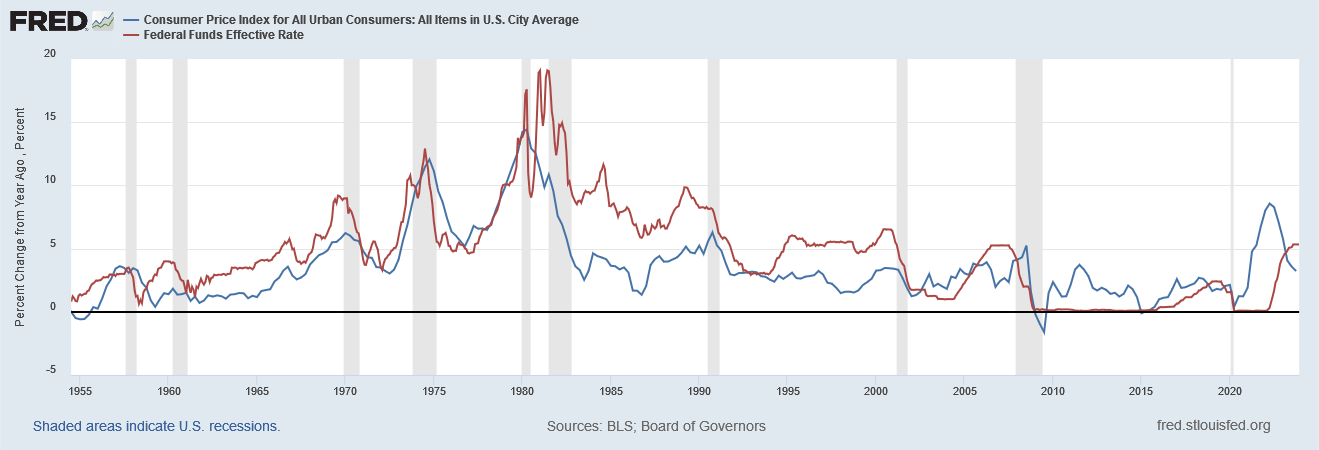}
\end{figure}
As shown in Figure \ref{fig:CPI_fundrate}, CPI YoY closely follows Fed Fund rate for the majority of the history. To quantify this close relation, we can calculate the correlation between them, which is 0.6860. We know sample correlation between non-stationary time series can be spurious, so we performed the ADF (augmented Dicky-Fuller) stationary test. Based on the $p$-value of 0.015 for CPI YoY and 0.10 for Federal funds rate, we can regard these two time series as stationary and take the 0.6860 correlation as a meaningful measure.  Actually besides Federal funds rates, we see CPI YoY has good correlations with various "spot" interest rates and/or their YoY growth, as listed in Table \ref{table:rates}, where DIFF12M, RDIFF12M and LDIFF12M represent respectively the 12-month simple difference, ratio difference and log difference. 

Besides the Federal funds rate and its 0.6860 correlation already mentioned above , we want to call out a list of relevant rates: 30-Year Primary Mortgage Rate level (0.5878) and DIFF12M (0.5523), 30-Year Jumbo Fixed Primary Mortgage Rate DIFF12M (0.6422), Cost of Funding Index (COFI) level (0.5557) and DIFF12M (0.5088), 1-Year Treasury Rate DIFF12M (0.6370). 

This strong relation is not surprising. Fisher equation (\cite{fisher1907rate}), a famous concept in the field of macroeconomics originated from economist Irving Fish, states that the nominal interest rate is equal to the sum of the real interest rate plus inflation. In equation, it can be represented as 
$$(1+i) = (1+r)\cdot (1+\pi) $$ which immediately gives 
$$i \approx r+\pi$$
where $i$ is the nominal interest rate, $\pi$ the expected inflation rate, and $r$ the real interest rate. According to the Fisher Equation, Fisher hypothesis asserts that the real interest rate is unaffacted by monetary policy and usually remains stable in short term. Therefore, with a fixed real interest rate, a given percent change in the expected inflation rate will necessarily be met with an equal percent change in teh nominal interest rate in the same direction. We see in Section \ref{sec:literature} that some literatures have used interest rates in their LGD models and all the interest rates (i.e., 3-month treasury bill, fed fund rate, etc.) they used are are nominal interest rates. Using CPI in LGD model can capture some of the effects of those nominal interest rates. 

Another interesting discovery is that, as shown in the last row of Table \ref{table:rates}, CPI YoY and UER has a very low correlation at 0.09. Given the UER from 1950 to 2022 can pass the stationary test, we think this suggests that CPI YoY and spot UER are nearly independent/orthogonal to each other and thus CPI YoY can be safely added to a GLM model where UER is present, with minimal change to the coefficient of UER. This is actually what we do observe in many of our models.

\subsection{CPI and future rates}
As already stressed several times, our CRE CTL defaults can take 12 or more months to foreclose and resolve, and thus compared to spot rates at defaults, the future rates 12 months after defaults are more relevant when determining LGD. To check if CPI YoY at time $t$ can provide some inference to rates after 12 months, in Table \ref{table:rates} we list the correlations between CPI YoY at time $t$ and 1) rates 12-months later or 2) rate changes during the following 12 months. For reader's convenience, we highlight 
\begin{itemize}
  \item in pink color the raw rates, which are generally non-stationary
  \item in yellow color the RDIFF12M, i.e., 12 months ratio difference
  \item in blue color the DIFF12M, i.e., 12 months simple difference which is a more nature and popular transformation compared to 12 months ratio difference
\end{itemize}

Again, the stationary Federal funds rate 1 year forward has a strong correlation with CPI YoY, at 0.6960, followed by 10-Year Treasury Rate (0.6205), 30-Year Treasury Rate (0.6091) and 7-Year Treasury Rate (0.4946). Note that, we see CPI YoY is also negatively correlated to S\&P500 RDIFF12M (-0.1077) and Dow Jones RDIFF12M (-0.2615). Recall in \ref{sec:lgd_rates} we mentioned \cite{betz2018systematic} found the US bank loans' LGD is negatively correlated to YoY log return of S\&P500 (i.e., the EI).

One thing worth mentioning is that, from 1952 to 2022, we see spot CPI YoY and CPI YoY 1 year forward have a correlation of 0.7254.  Also, CPI YoY has a moderate 0.32 correlation to UER 1 year forward. Recall in Section 2, we see many authors find UER to be positively correlated to LGD (see \ref{table:literature}).

\subsection{CPI and future CRE market sales price}
We already presented in above section the (strong) positive correlation between CPI YoY and various spot/future rates. And we argue a high CPI YoY at time $t$ implies high rates at time $t$ and  the following 1 year window, and thus lowers CRE properties' value and sale price. Actually, we can verify this claim in a more straightforward way. 

Let us denote as $B$, $V$ and $W$ the loan balance, property value and workout cost, and use subscript $d$ and $s$ to label these quantities at default and sale. By the definition of workout LGD, and assuming $B_d = B_s$, i.e., loan balance does not change between default and sale, we have 
\begin{equation*}
\text{LGD} = \frac{B_{d} + W_{d,s} - V_{s}}{B_d} = 1 - \frac{V_s}{B_d}  + \frac{W_{d,s}}{B_d} = 1 - \frac{1}{\text{LTV}_{s}} + \frac{W_{d,s} }{B_d}.
\end{equation*}
In the meanwhile, it holds that
\begin{equation*}
\text{LTV}_{s} \triangleq \frac{B_{s}}{V_{s}} = \frac{B_{d}}{V_{d}}\cdot  \frac{V_{d}}{V_{s}}= \text{LTV}_{d}  \cdot \frac{V_{d}}{V_{s}}
\end{equation*}
Combining the above two equations,  with $\Delta V_{d,s} \triangleq V_s - V_d$, we get 
\begin{align}
\text{LGD} &=  1 - \frac{V_s}{V_d}\cdot\frac{1}{\text{LTV}_{d}} + \frac{W_{d,s}}{B_d}  \nonumber \\ 
                 &= 1 -  (1 + \frac{\Delta V_{d,s}}{V_d})\cdot\frac{1}{\text{LTV}_{d}} + \frac{W_{d,s}}{B_d}  \nonumber \\ 
                 &= 1 - \frac{1}{\text{LTV}_{d}} -   \frac{\Delta V_{d,s}}{V_d}\cdot\frac{1}{\text{LTV}_{d}} + \frac{W_{d,s}}{B_d}  
\label{eq: LGD1}
\end{align}
$\Delta V_{d,s}/V_d$ can be further break down into market price change between $d$ and $s$, loan level factors (such as location) and hard to catch idiosyncrasy. That is, we can write 
\begin{equation}
\label{eq: LGD2}
 \frac{\Delta V_{d,s}}{V_d} = \text{CRE Market Price Change}_{d,s}+ \text{Loan level factors} + \epsilon 
\end{equation}

Assuming $\text{LTV}_d$, $V_d$ and $B_d$ are already fixed at the time of default, a higher CPI YoY at defaulting time $d$ implies higher CPI YoY one year later (with correlation 0.7254), and can raise $W_{d,s}$, the workout cost, which includes all fees and costs related to legal, collection and property maintenance. The relation between CPI YoY at time $d$ and $\Delta V_{d,s}/V_d$ can in certain way revealed by our CRE CTL LGD data.

\subsubsection{With real resolution time}
Our CRE CTL LGD data consists of about 4000 resolved defaults. For every of the 4000 entries, we can get a pair of $d$ and $s$, and the resolution time in month, $s-d$, varies from 0 to 120 months.  The correlation calculation procedure is outlined below:
\begin{itemize}
  \item Select defaults having an appraisal within 6 months of default dates 
  \item Use each of these defaults as a data point, and calculate the individual $\Delta V_{d,s}/V_d$
  \item Calculate the correlation between $\Delta V_{d,s}/V_d$ and CPI YoY at $d$, across all defaults
\end{itemize}
Following this approach, we get a correlation at \textbf{-0.3352} for CPI YoY at $d$ and $\Delta V_{d,s}/V_d$.

However, there are several issues associated this approach. 

Firstly, in our data it is hard to identify the exact and accurate information about the final sale price, which is often combined with workout cost and other adjustments. 

Secondly, by using CRE CTL LGD data from JPMC, the calculated correlation may be biased towards JPMC's LGD and/or loan patterns and not transferable to other portfolios. Further, in a given window the correlation can be distorted by the loan level factors mentioned in equation \ref{eq: LGD2} in that window and thus not reflecting the underlying truth. 

Lastly, given our defaults are concentrated in GFC, this default level correlation is inevitably dominated by the short term correlation  pattern in GFC and thus not representative.

\subsubsection{With a flat 12 months resolution time}
\label{sec:cpi_value}
As an improved approach, we now aim to get the correlation between CPI YoY at $d$ and CRE market price change between $d$ and $s$. We assume a flat 12 months resolution time, i.e., letting $s = d + 12$. Meanwhile, to better capture the CRE market price change, instead of the US national level CREPI (commercial real estate price index) and the property type level CCRSI (CoStar commercial repeat sales indices), we use the more granular monthly CoStar market sales price indices that cover US and 54 key MSAs and 4 property types (Apartment/Industry/Office/Retail). So there are $55 \times 4 = 220$ sales price indices, and each of them is available back to 1982, and its YoY starts at 1983. This enables us to calculate the correlation for any window of interest from 1983 to 2022. 

\begin{figure}[h!]
	\caption{Diagram of Calculating the Correlation between CPI and CRE Market Sales Price Indices}
         \label{fig:two_approach_diagram}
	\centering
	\includegraphics[scale=0.8]{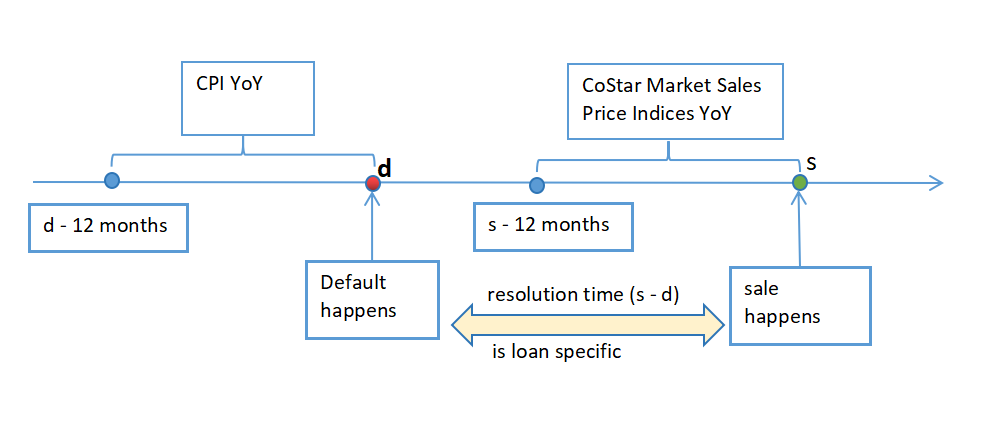}
\end{figure}

Given our CRE CTL portfolio mainly consists of apartment, and the current worry over office sector, we check apartment and office market sales price indices, at national and key markets like New York and LA, during five windows of interest in history. The result is listed in Table \ref{table: Costar_CPI}. We do see consistent negative correlations between CPI YoY and APT market price indices YoY 1 year later in various windows, such as the all year window from 1983 to 2022, GFC, early 1990s recession and COVID-19. In GFC, they can go above -0.80; in COVID-19 and 1990 recession, they are at high -0.50. For office, we also see good negative correlations in all windows, except for COVID-19 period where we know there could be some pattern change for office sector.

\subsubsection{By different CPI YoY buckets}
We also checked the above correlations by different CPI YoY buckets, i.e., $<= 2\%$, $2\%\sim4\%$ and $>4\%$. The results are listed in Table \ref{table: Costar_CPI_byBucket}, with Table \ref{table: Costar_CPI_high} being the correlations at high CPI YoY bucket (i.e., $> 4\%$) break down by different historical windows. 

The observation is that, in the time when CPI YoY is in $2\%\sim4\%$ bucket, it still has consistent negative correlations with apartment and office market sales price indices YoY 1 year forward, at national, New York and LA. 

As shown in Table \ref{table: Costar_CPI_high}, when CPI YoY is in $>4\%$ bucket, CPI YoY shows very weak positive correlation with CRE market sales price YoY q year forward at National APT (0.1233) and LA APT (0.0544).
It shows stronger than usual negative correlations to those indices YoY 1 year forward in GFC and COVID-19 period, with correlations reach -45\% in GFC and -70\% or even -90\% in COVID-19. 

These negative correlations between CPI YoY and various CRE market sales price indices, together with the "-" sign in front of  $\Delta V_{d,s}/V_d$ in equation (\ref{eq: LGD1}), can ensure a positive correlation between  CPI YoY and the spot LGD.  
\begin{table}[h]
\centering
\caption{Correlation between CPI YoY and CoStar Market Sales Price Index YoY 1 Year Later, monthly data, for CPI YoY $>$ 0.04 Bucket, Breakdown by Historical Windows }
\label{table: Costar_CPI_high}
\scalebox{0.65}{
\begin{tabular}{|c|c|l|l|r|}
\hline
\textbf{CPI YoY Bucket}             & \textbf{Count}       & \textbf{MSA, Prop}             & \textbf{Window}           & \textbf{Correlation} \\ \hline
\multirow{24}{*}{\textgreater 0.04} & \multirow{24}{*}{85} & \multirow{4}{*}{US National, APT}       & Before 2004 (1980s \& 1990s) & 0.1233               \\ \cline{4-5} 
                                    &                      &                                & 2004-2019 (modeling period)  & -0.1821              \\ \cline{4-5} 
                                    &                      &                                & 2007-2010 (GFC)              & -0.2479              \\ \cline{4-5} 
                                    &                      &                                & 2020-2022 (Covid)            & -0.9325              \\ \cline{3-5} 
                                    &                      & \multirow{4}{*}{US National, OFF}       & Before 2004 (1980s \& 1990s) & -0.2727              \\ \cline{4-5} 
                                    &                      &                                & 2004-2019 (modeling period)  & -0.3676              \\ \cline{4-5} 
                                    &                      &                                & 2007-2010 (GFC)              & -0.4632              \\ \cline{4-5} 
                                    &                      &                                & 2020-2022 (Covid)            & -0.7313              \\ \cline{3-5} 
                                    &                      & \multirow{4}{*}{New York, APT} & Before 2004 (1980s \& 1990s) & -0.1393              \\ \cline{4-5} 
                                    &                      &                                & 2004-2019 (modeling period)  & -0.2275              \\ \cline{4-5} 
                                    &                      &                                & 2007-2010 (GFC)              & -0.2992              \\ \cline{4-5} 
                                    &                      &                                & 2020-2022 (Covid)            & -0.9664              \\ \cline{3-5} 
                                    &                      & \multirow{4}{*}{New York, OFF} & Before 2004 (1980s \& 1990s) & -0.4564          \\ \cline{4-5} 
                                    &                      &                                & 2004-2019 (modeling period)  & -0.4067            \\ \cline{4-5} 
                                    &                      &                                & 2007-2010 (GFC)              & -0.4959              \\ \cline{4-5} 
                                    &                      &                                & 2020-2022 (Covid)            & -0.9830             \\ \cline{3-5} 
                                    &                      & \multirow{4}{*}{LA, APT}       & Before 2004 (1980s \& 1990s) & 0.0544              \\ \cline{4-5} 
                                    &                      &                                & 2004-2019 (modeling period)  & -0.2340              \\ \cline{4-5} 
                                    &                      &                                & 2007-2010 (GFC)              & -0.3174              \\ \cline{4-5} 
                                    &                      &                                & 2020-2022 (Covid)            & -0.9147              \\ \cline{3-5} 
                                    &                      & \multirow{4}{*}{LA, OFF}       & Before 2004 (1980s \& 1990s) & -0.3005              \\ \cline{4-5} 
                                    &                      &                                & 2004-2019 (modeling period)  & -0.3699                \\ \cline{4-5} 
                                    &                      &                                & 2007-2010 (GFC)              & -0.4656                \\ \cline{4-5} 
                                    &                      &                                & 2020-2022 (Covid)            & -0.8932                 \\ \hline
\end{tabular}
}
\end{table}

\subsection{Economic Meaning related to deferred maintenances and tenant improvement}
Besides pure analyses, we have also consulted expertise from CRE SMEs who also support using CPI in LGD because of the deferred maintenance. In CRE industry, banks generally only perform minimum or even no maintenance (lawn, roof, exterior wall, etc.) to foreclosed properties from defaulted loans. Potential buyers will adjust their bids to price in the cost for these deferred maintenances on their side, considering all due repairs and renovations. These costs are obviously subject to inflation. Thus under high inflation scenario, potential buyers are more likely to submit lower bids because of higher maintenance cost they anticipate to pay once get the property. The lower bids can lead to lower recovery and higher LGD. This is the "workout cost" on the buyer side which priced into recovery we can receive. 

Business also told us that there will be tenant improvement (TI) money, which represents that for the sold/foreclosed property, the new owner will need to invest money into the building restructure and/or improvements or offering discount deals in order to maintain the current tenants in the building. When inflation is high, this cost will be high, which is also priced into the bid the potential owner will offer. This TI cost is another "workout cost" on the buyer side. Furthermore, the current insurance costs for CRE building is running high due to the inflation and climate risks, which is another headache to the potential buyers. 

Not only the deferred maintenance and TI, when CPI is high, the salary for hiring operation personals, lawyers, etc., will also increase, leads to increasing "workout cost" on our side.

\section{LGD Model with CPI}
In Section \ref{sec:cpi_rate} we presented the correlation between CPI YoY, various spot rates and future rates, and CRE market sales price indices, which suggest CPI can provide inference to CRE LGD. With the CRE CTL LGD data from JPMC consists of 4000 resolved defaults from 2004-01 to 2019-12,  we built a Tobit I model to forecast $\text{LGD}^+ = \text{LGD}\cdot I(\text{LGD}>=0)$, i.e., the LGD left-censored at 0. That is, we assume for the $i$th default,
\begin{equation*}
\text{LGD}_i = x^T_i\beta + \epsilon_i, \qquad \epsilon_i|x_i \stackrel{iid}{\sim} N(0,\sigma^2).
\end{equation*}
where $x_i$ is the relevant loan attributes and macroeconomic variables observed no later than the loan's defaulting time.
Thus we have
\begin{align*}
E(\text{LGD}_i^+|x_i) &= E(\text{LGD}_i|\text{LGD}_i>0, x_i)\cdot P(\text{LGD}_i>0|x_i)   \\
                            &=\Phi(x^T_i\beta)\big( x^T_i\beta + \sigma \cdot \frac{\phi(x^T_i\beta) }{\Phi(x^T_i\beta)}\big)
\end{align*}
where $\phi(\cdot)$ and $\Phi(\cdot)$ are the density and CDF of standard normal distribution. With known $x_i$, the coefficients $\beta$ can be estimated using MLE after a reparametrization, as mentioned by \cite{olsen1978note}. 

\subsection{Estimates}

The detailed variable selection algorithm to determine relevant $x_i$, especially from a long list of MEVs and their transformations is not the main topic and thus skipped here. Readers interested in this topic can refer to our paper \cite{arora2024exploration} for methodology pertinent to MEV screen and selection.  . 

We found various HPI and CPI transformations, together with other loan level attributes like LTV are significant in our Tobit I LGD model, while the model with HPI and CPI LDIFF6M, i.e., 6 months log difference, are giving the best (lowest) BIC (Bayes information criterion, see e.g., Section 7.7 in \cite{hastie2009elements}). 

\begin{figure}[h!]
	\caption{LGD model in-sample Fitness}
         \label{fig:in_sample_fit}
	\centering
	\includegraphics[scale=0.6]{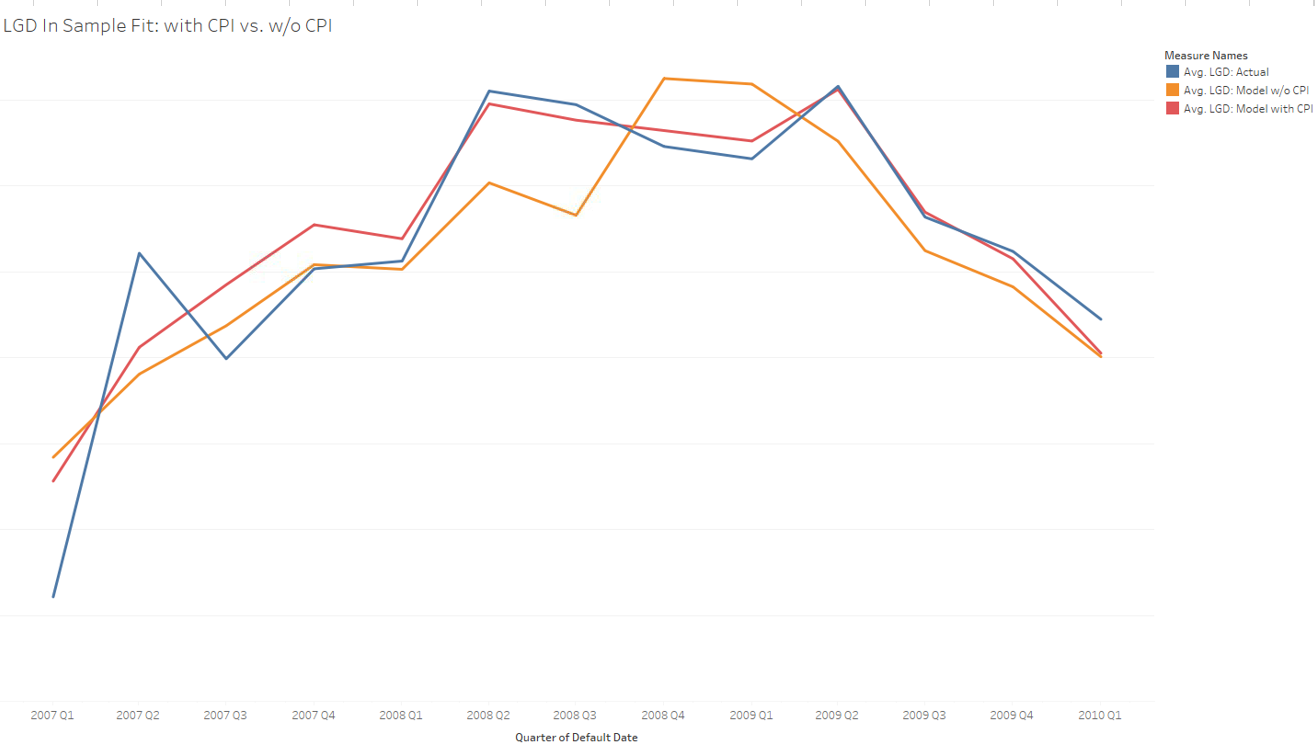}
\end{figure}

The coefficients estimate and $t$-test for significance are listed in Table \ref{table:lgd_coeff} (other covariates are redacted to protect JPMC internal information). In Section \ref{sec:cpi_rate}, we already presented various evidences that suggest a positive correlation between CPI YoY and LGD. Here the 2.40 coefficient of CPI 6 month log difference is expected. 
\begin{table}[h]
\center
\caption{LGD Champion Model Coefficients, Redacted Version}\label{table:lgd_coeff}
\begin{tabular}{|lrrrr|}
\hline
\multicolumn{1}{|l|}{\textbf{Covariate}} & \multicolumn{1}{l|}{\textbf{Estimate}} & \multicolumn{1}{l|}{\textbf{\begin{tabular}[c]{@{}l@{}}Standard \\ Error\end{tabular}}} & \multicolumn{1}{l|}{\textbf{$z$\_value}} & \multicolumn{1}{l|}{\textbf{\begin{tabular}[c]{@{}l@{}}Approx. \\ $P >|z|$\end{tabular}}} \\ \hline
\multicolumn{1}{|l|}{HPI\_M.LDIFF6M}     & \multicolumn{1}{r|}{-2.23}             & \multicolumn{1}{r|}{0.370}                                                              & \multicolumn{1}{r|}{-6.037}            & 0.000                                                                                               \\ \hline
\multicolumn{1}{|l|}{CPI\_M.LDIFF6M}     & \multicolumn{1}{r|}{2.40}              & \multicolumn{1}{r|}{0.496}                                                              & \multicolumn{1}{r|}{4.842}             & 0.000                                                                                               \\ \hline
\multicolumn{5}{|c|}{\textbf{Other Covariates Redacted}}                                                                                                                                                                                                                                                                   \\ \hline
\end{tabular}
\end{table}
To see how CPI helps to address the early downturn LGD underestimation issue, we plot in Figure \ref{fig:in_sample_fit} the quarterly in-sample simple average LGD, actual (blue) vs. champion (with CPI, red) vs. best model without CPI. The "best model without CPI" here refers to the corresponding model with smallest BIC when we remove CPI from the candidate variable pool. It is clear that the champion model with CPI can capture the actual LGD peak in 2008 and 2009. Note that, to protect JPMC internal information, we hide the Y axis values and the LGD curves outside of the 2007-2010 window.

We understand that BIC, as a model assessment and selection criterion, may not select the model that optimizes the downturn fitness. We actually calculate for all candidate models the underestimation at 2008Q2 and then sort from low to high. The champion with CPI ranks the best with the lowest underestimation, while models without CPI shows consistent 2008Q2 underestimation. We think this suggests the CPI's usefulness in capturing early downturn LGD peak, at least for our data. 

\subsection{Cross validation results}
To prevent overfitting and spurious regression result between CPI and LGD, we further perform two types of cross validation to check the stability of the estimated coefficient of CPI.

The first is the well known 10-fold cross validation. Besides the estimated out-of-sample prediction error, what we care more is the stability of the coefficient of the CPI predictor, since an instable coefficient is a strong indicator of overfitting. Table \ref{table:10-fold} summarizes the result of such test. When using the whole data, we get coefficient 2.40 and -2.23 for CPI and HPI predictor respectively; when the 1st fold is held out and the model is refitted using the rest 9 folds, the coefficients become 2.46 and -2.30 respectively. We see the 10 refitted coefficients, for both HPI and CPI, are tightly distributed and all within 1 std of that fitted with the whole data.
\begin{table}[h]
\center
\caption{10-Fold Cross Validation on Coefficient Stability}\label{table:10-fold}
\scalebox{0.56}{
\begin{tabular}{lrrrrrrrrrrrr}
\cline{2-13}
\multicolumn{1}{l|}{}                         & \multicolumn{2}{c|}{\textbf{All Data}}                                 & \multicolumn{10}{c|}{\textbf{10-Fold Cross Validation}}                                                                                                                                                                                                                                                                                                                                              \\ \cline{2-13} 
\multicolumn{1}{l|}{}                         & \multicolumn{1}{l|}{\textbf{Est.}} & \multicolumn{1}{l|}{\textbf{S.E}} & \multicolumn{1}{l|}{\textbf{Fold 1}} & \multicolumn{1}{l|}{\textbf{Fold 2}} & \multicolumn{1}{l|}{\textbf{Fold 3}} & \multicolumn{1}{l|}{\textbf{Fold 4}} & \multicolumn{1}{l|}{\textbf{Fold 5}} & \multicolumn{1}{l|}{\textbf{Fold 6}} & \multicolumn{1}{l|}{\textbf{Fold 7}} & \multicolumn{1}{l|}{\textbf{Fold 8}} & \multicolumn{1}{l|}{\textbf{Fold 9}} & \multicolumn{1}{l|}{\textbf{Fold 10}} \\ \hline
\multicolumn{1}{|l|}{\textbf{HPI\_M.LDIFF6M}} & \multicolumn{1}{r|}{-2.23}         & \multicolumn{1}{r|}{0.37}         & \multicolumn{1}{r|}{-2.30}           & \multicolumn{1}{r|}{-2.15}           & \multicolumn{1}{r|}{-2.25}           & \multicolumn{1}{r|}{-2.29}           & \multicolumn{1}{r|}{-2.26}           & \multicolumn{1}{r|}{-2.25}           & \multicolumn{1}{r|}{-2.14}           & \multicolumn{1}{r|}{-2.27}           & \multicolumn{1}{r|}{-2.12}           & \multicolumn{1}{r|}{-2.28}            \\ \hline
\multicolumn{1}{|l|}{\textbf{CPI\_M.LDIFF6M}} & \multicolumn{1}{r|}{2.40}          & \multicolumn{1}{r|}{0.50}         & \multicolumn{1}{r|}{2.46}            & \multicolumn{1}{r|}{2.26}            & \multicolumn{1}{r|}{2.68}            & \multicolumn{1}{r|}{2.41}            & \multicolumn{1}{r|}{2.59}            & \multicolumn{1}{r|}{2.23}            & \multicolumn{1}{r|}{2.24}            & \multicolumn{1}{r|}{2.41}            & \multicolumn{1}{r|}{2.49}            & \multicolumn{1}{r|}{2.25}             \\ \hline
\multicolumn{13}{|c|}{\textbf{Other Covariates Redacted}}                                                                                                                                                                                                                                                                                                                                                                                                                                                                             \\ \hline
\end{tabular}
}
\end{table}

The second is the "leave one year out" cross validation. Given the CRE CTL LGD data spans from 2004 to 2019, it is possible that there are several distinct patterns. To detect that, in each iteration we hold out one year's data and refit the model using the data from the rest years. The test result is presented in Table \ref{table:leave_one_year}. We notice that, with 2009 data (containing more than 25\% of the all years total 4000 defaults) held out, the coefficient of CPI is lowered from 2.40 to 1.74. This is consistent to what we observe in Section \ref{sec:cpi_value}, Table \ref{table: Costar_CPI} and \ref{table: Costar_CPI_byBucket}: During GFC, CPI YoY exhibited much higher negative correlation with the future (1 year later) CRE market sales price indices YoY, which can lead to much higher positive correlation and coefficient to LGD. And this explains why removing 2009 data lowers CPI coefficient in our model. Given one essential application of our LGD model is to predict CRE CTL LGD and losses in severe economy scenarios,  we actually prefer our model to learn more from the LGD pattern in GFC. Also, the 1.74 coefficient is still within the 2std range of 2.40 and cannot be regarded as a break. 
\begin{table}[h]
\center
\caption{Leave One Year Out Cross Validation on Coefficient Stability}\label{table:leave_one_year}
\scalebox{0.47}{
\begin{tabular}{lcrcccccccccccccccc|}
\cline{2-19}
\multicolumn{1}{l|}{}                         & \multicolumn{2}{c|}{\textbf{All Year}}                                  & \multicolumn{16}{c|}{\textbf{Leave One Year Out Cross Validation}}                                                                                                                                                                                                                                                                                                                                                                                                                                                                                                                                    \\ \cline{2-19} 
\multicolumn{1}{l|}{}                         & \multicolumn{1}{c|}{\textbf{Est.}} & \multicolumn{1}{c|}{\textbf{S.E.}} & \multicolumn{1}{c|}{\textbf{2004}} & \multicolumn{1}{c|}{\textbf{2005}} & \multicolumn{1}{c|}{\textbf{2006}} & \multicolumn{1}{c|}{\textbf{2007}} & \multicolumn{1}{c|}{\textbf{2008}} & \multicolumn{1}{c|}{\textbf{2009}} & \multicolumn{1}{c|}{\textbf{2010}} & \multicolumn{1}{c|}{\textbf{2011}} & \multicolumn{1}{c|}{\textbf{2012}} & \multicolumn{1}{c|}{\textbf{2013}} & \multicolumn{1}{c|}{\textbf{2014}} & \multicolumn{1}{c|}{\textbf{2015}} & \multicolumn{1}{c|}{\textbf{2016}} & \multicolumn{1}{c|}{\textbf{2017}} & \multicolumn{1}{c|}{\textbf{2018}} & \textbf{2019}              \\ \hline
\multicolumn{1}{|l|}{\textbf{HPI\_M.LDIFF6M}} & \multicolumn{1}{r|}{-2.23}         & \multicolumn{1}{r|}{0.37}          & \multicolumn{1}{r|}{-2.11}         & \multicolumn{1}{r|}{-2.14}         & \multicolumn{1}{r|}{-2.04}         & \multicolumn{1}{r|}{-1.98}         & \multicolumn{1}{r|}{-2.14}         & \multicolumn{1}{r|}{-3.37}         & \multicolumn{1}{r|}{-2.42}         & \multicolumn{1}{r|}{-2.11}         & \multicolumn{1}{r|}{-2.39}         & \multicolumn{1}{r|}{-2.32}         & \multicolumn{1}{r|}{-2.21}         & \multicolumn{1}{r|}{-2.13}         & \multicolumn{1}{r|}{-2.16}         & \multicolumn{1}{r|}{-2.18}         & \multicolumn{1}{r|}{-2.19}         & \multicolumn{1}{r|}{-2.22} \\ \hline
\multicolumn{1}{|l|}{\textbf{CPI\_M.LDIFF6M}} & \multicolumn{1}{r|}{2.40}          & \multicolumn{1}{r|}{0.50}          & \multicolumn{1}{r|}{2.34}          & \multicolumn{1}{r|}{2.40}          & \multicolumn{1}{r|}{2.35}          & \multicolumn{1}{r|}{2.34}          & \multicolumn{1}{r|}{2.59}          & \multicolumn{1}{r|}{1.74}          & \multicolumn{1}{r|}{2.71}          & \multicolumn{1}{r|}{2.24}          & \multicolumn{1}{r|}{2.62}          & \multicolumn{1}{r|}{2.69}          & \multicolumn{1}{r|}{2.33}          & \multicolumn{1}{r|}{2.22}          & \multicolumn{1}{r|}{2.30}          & \multicolumn{1}{r|}{2.33}          & \multicolumn{1}{r|}{2.41}          & \multicolumn{1}{r|}{2.39}  \\ \hline
\multicolumn{19}{|c|}{\textbf{Other Covariates Redacted}}                                                                                                                                                                                                                                                                                                                                                                                                                                                                                                                                                                                                                                                                       \\ \hline
\multicolumn{1}{|l|}{\textbf{\# of Samples}}  & \multicolumn{1}{l|}{}              & \multicolumn{1}{l|}{}              & \multicolumn{1}{c|}{52}            & \multicolumn{1}{c|}{51}            & \multicolumn{1}{c|}{77}            & \multicolumn{1}{c|}{171}           & \multicolumn{1}{c|}{353}           & \multicolumn{1}{c|}{1,160}         & \multicolumn{1}{c|}{879}           & \multicolumn{1}{c|}{459}           & \multicolumn{1}{c|}{250}           & \multicolumn{1}{c|}{188}           & \multicolumn{1}{c|}{104}           & \multicolumn{1}{c|}{91}            & \multicolumn{1}{c|}{98}            & \multicolumn{1}{c|}{59}            & \multicolumn{1}{c|}{63}            & 9                          \\ \hline
\end{tabular}
}
\end{table}

\subsection{Validation for nonlinear patterns}
When we presented this model to the Model Review and Governance Group within JPMC, we received a concern over the use of CPI LDIFF6M in LGD model with a positive coefficient. Their worry is that, though this CPI predictor can increase LGD forecast during stagflation scenario with a stemming CPI, it may also cause a lower LGD in stress scenarios where CPI pummels (i.e., negative CPI LDIFF6M) because the high unemployment and crashing economy may lead to much reduced demand. CPI LDIFF6M could have a "V" shape relation with LGD and our Tobit I model, which essentially is a linear regression model, may be over-simplified when using the CPI LDIFF6M term.  

We think this is a very reasonable challenge. And to detect potential "V" shape or more complicated patterns between CPI and LGD, we enlist the multivariate adaptive regression splines (MARS) introduced by \cite{friedman1991multivariate}. Basically, it is a powerful machine learning algorithm which can automatically detect nonlinear patterns and interactions between variables. For a quick yet detailed introduction of this algorithm, besides the original paper one can refer to Section 9.4 in \cite{hastie2009elements}.

\begin{figure}[h]
	\caption{Quadratic Pattern between CPI LDIFF6M and LGD, Detected by MARS}
         \label{fig:MARS}
	\centering
	\includegraphics[scale=0.6]{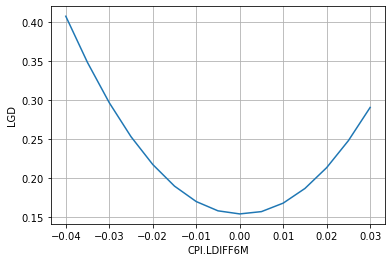}
\end{figure}

With LGD as the target and CPI LDIFF6M as the only predictor, MARS asserts a quadratic + hinge function pattern between CPI LDIFF6M and our LGD, as plotted in Figure \ref{fig:MARS}, with the specification given in Table \ref{table:mars}. 

\begin{table}[h]
\center
\caption{Quadratic Pattern Specification}\label{table:mars}
\begin{tabular}{|l|l|r|}
\hline
  & \multicolumn{1}{c|}{\textbf{Term}} & \multicolumn{1}{c|}{\textbf{Coefficient}} \\ \hline
1 & (Intercept)                        & 0.1541                                    \\ \hline
2 & CPI\_M.LDIFF6M * CPI\_M.LDIFF6M    & 158.526                                   \\ \hline
3 & max(0, CPI\_M.LDIFF6M - 0)           & -0.20088                                  \\ \hline
\end{tabular}
\end{table}

\begin{figure}[h]
	\caption{MARS Result with Both HPI and CPI LDIFF6M as Predictors}
         \label{fig:mars}
	\centering
	\includegraphics[scale=0.6]{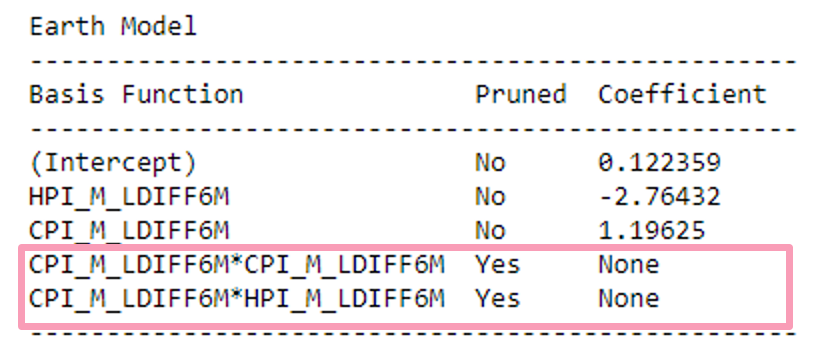}
\end{figure}

However, once we add in the second predictor HPI LDIFF6M, MARS regards the quadratic term of CPI LIDFF6M and the interaction between HPI and CPI redundant and decides to prune them, as shown in Figure \ref{fig:mars}.

This suggest that, when using CPI LDIFF6M in tandem with HPI LDIFF6M, their simple linear term can already capture the LGD and there is no need of piecewise (hinge function),  quadratic or interaction terms. The high LGD observed at very negative CPI LDIFF6M can be well explained by negative HPI LDIFF6M so MARS decides to drop the "V" shape pattern of CPI LDIFF6M.

\section{Conclusion}
In this paper, we explained why forecasting CRE LGD using MEVs only up to default time is challenging given CRE defaulted loans' long workout and foreclosure period, and why CPI can contribute to solve this difficulty. We presented the strong positive correlation between CPI YoY and 1) various spot rates and rates transformations 2) various future rates and rates transformations 3) CRE market sales price indices in different historical windows from 1982 to 2022, including GFC, COVID-19, etc., for both apartment and office properties in New York,  LA and US national, also at different CPI YoY buckets. We further provided economic and business intuitions and supports. At last, we developed a Tobit I LGD model using CRE CTL data that consists of 4000 defaults from 2004 to 2019, with positive and significant CPI 6 months log difference as a predictor, which can capture the early GFC LGD peak. Various cross validations clear this model from spurious regression and the MARS algorithm shows that the linear term of HPI and CPI 6 month log difference is suffice to capture LGD and no nonlinear patterns such as piecewise spline and quadratic terms are needed. Based on all these, we believe CPI is a good leading indicator to CRE market and can improve the accuracy of LGD models for CRE portfolio. 

As an end, we want to stress the observational study nature of all our work presented in this paper, as we do not have the ability to conduct experiments to CPI. Yet given the nowadays persistent inflation and the worry of a upcoming stagflation,  we suggest modelers and researches to at least consider and test CPI when developing their CRE LGD forecasting models. 

\section*{CRediT authorship contribution statement}
\textbf{Ying Wu}: Methodology, Resources, Software, Formal analysis, Investigation, Validation, Data curation, Writing - original draft.
\textbf{Garvit Arora}: Methodology, Software, Formal analysis,  Investigation. 
\textbf{Xuan Mei}: Conceptualization, Methodology, Software, Formal analysis, Investigation, Validation, Data curation, Resources, Writing - original draft, Writing - review \& editing, Project administration. 

\section*{Acknowledgements}
We gratefully acknowledge the support and encouragement we received from Junze Lin and Stuart Marker for our research on this topic. Some of the analyses are suggested by Ruiqi Zhang and Jenny Wang. We are also deeply indebted to Bjoern Hinrichsen for his heartful suggestions and discussions that drastically improved this paper and made it compliant to JPMC external
publication polices.

\bibliographystyle{elsarticle-harv}
\bibliography{reference}

\begin{thebibliography}{20}
\expandafter\ifx\csname natexlab\endcsname\relax\def\natexlab#1{#1}\fi
\providecommand{\url}[1]{\texttt{#1}}
\providecommand{\href}[2]{#2}
\providecommand{\path}[1]{#1}
\providecommand{\DOIprefix}{doi:}
\providecommand{\ArXivprefix}{arXiv:}
\providecommand{\URLprefix}{URL: }
\providecommand{\Pubmedprefix}{pmid:}
\providecommand{\doi}[1]{\href{http://dx.doi.org/#1}{\path{#1}}}
\providecommand{\Pubmed}[1]{\href{pmid:#1}{\path{#1}}}
\providecommand{\bibinfo}[2]{#2}
\ifx\xfnm\relax \def\xfnm[#1]{\unskip,\space#1}\fi
\bibitem[{Arora et~al.(2024)Arora, Shubhangi, Wu and
  Mei}]{arora2024exploration}
\bibinfo{author}{Arora, G.}, \bibinfo{author}{Shubhangi, S.},
  \bibinfo{author}{Wu, Y.}, \bibinfo{author}{Mei, X.}, \bibinfo{year}{2024}.
\newblock \bibinfo{title}{An exploration to the correlation structure and
  clustering of macroeconomic variables (mev)}.
\newblock \bibinfo{journal}{arXiv preprint arXiv:2401.10162} .
\bibitem[{Bellotti and Crook(2012)}]{bellotti2012loss}
\bibinfo{author}{Bellotti, T.}, \bibinfo{author}{Crook, J.},
  \bibinfo{year}{2012}.
\newblock \bibinfo{title}{Loss given default models incorporating macroeconomic
  variables for credit cards}.
\newblock \bibinfo{journal}{International Journal of Forecasting}
  \bibinfo{volume}{28}, \bibinfo{pages}{171--182}.
\bibitem[{Betz et~al.(2018)Betz, Kellner and R{\"o}sch}]{betz2018systematic}
\bibinfo{author}{Betz, J.}, \bibinfo{author}{Kellner, R.},
  \bibinfo{author}{R{\"o}sch, D.}, \bibinfo{year}{2018}.
\newblock \bibinfo{title}{Systematic effects among loss given defaults and
  their implications on downturn estimation}.
\newblock \bibinfo{journal}{European Journal of Operational Research}
  \bibinfo{volume}{271}, \bibinfo{pages}{1113--1144}.
\bibitem[{CoStar(2023)}]{CoStar_CCRSI}
\bibinfo{author}{CoStar}, \bibinfo{year}{2023}.
\newblock \bibinfo{title}{Costar commerical repeat-sale indices press release}.
\newblock \URLprefix
  \url{https://www.costargroup.com/sites/costargroup.com/files/docs/2024-01/CCRSI\_Jan2024.pdf}.
\bibitem[{Fisher and Barber(1907)}]{fisher1907rate}
\bibinfo{author}{Fisher, I.}, \bibinfo{author}{Barber, W.J.},
  \bibinfo{year}{1907}.
\newblock \bibinfo{title}{The rate of interest}.
\newblock \bibinfo{publisher}{Garland Pub.}
\bibitem[{FRB(2016)}]{CECL2016}
\bibinfo{author}{FRB}, \bibinfo{year}{2016}.
\newblock \bibinfo{title}{Financial instruments-credit losses (topic 326)}.
\newblock \URLprefix
  \url{https://www.fasb.org/Page/ShowPdf?path=ASU+2016-13.pdf}.
\bibitem[{FRB(2020)}]{CCAR2020}
\bibinfo{author}{FRB}, \bibinfo{year}{2020}.
\newblock \bibinfo{title}{Comprehensive capital analysis and review 2020
  summary instructions}.
\newblock \URLprefix
  \url{https://www.federalreserve.gov/newsevents/pressreleases/files/bcreg20200304a3.pdf}.
\bibitem[{Friedman(1991)}]{friedman1991multivariate}
\bibinfo{author}{Friedman, J.H.}, \bibinfo{year}{1991}.
\newblock \bibinfo{title}{Multivariate adaptive regression splines}.
\newblock \bibinfo{journal}{The annals of statistics} \bibinfo{volume}{19},
  \bibinfo{pages}{1--67}.
\bibitem[{Hastie et~al.(2009)Hastie, Tibshirani, Friedman and
  Friedman}]{hastie2009elements}
\bibinfo{author}{Hastie, T.}, \bibinfo{author}{Tibshirani, R.},
  \bibinfo{author}{Friedman, J.H.}, \bibinfo{author}{Friedman, J.H.},
  \bibinfo{year}{2009}.
\newblock \bibinfo{title}{The elements of statistical learning: data mining,
  inference, and prediction}. volume~\bibinfo{volume}{2}.
\newblock \bibinfo{publisher}{Springer}.
\bibitem[{Heckman(1979)}]{heckman1979sample}
\bibinfo{author}{Heckman, J.J.}, \bibinfo{year}{1979}.
\newblock \bibinfo{title}{Sample selection bias as a specification error}.
\newblock \bibinfo{journal}{Econometrica: Journal of the econometric society} ,
  \bibinfo{pages}{153--161}.
\bibitem[{Jankowitsch et~al.(2014)Jankowitsch, Nagler and
  Subrahmanyam}]{jankowitsch2014determinants}
\bibinfo{author}{Jankowitsch, R.}, \bibinfo{author}{Nagler, F.},
  \bibinfo{author}{Subrahmanyam, M.G.}, \bibinfo{year}{2014}.
\newblock \bibinfo{title}{The determinants of recovery rates in the us
  corporate bond market}.
\newblock \bibinfo{journal}{Journal of Financial Economics}
  \bibinfo{volume}{114}, \bibinfo{pages}{155--177}.
\bibitem[{Li et~al.(2023)Li, Li and Bellotti}]{li2023predicting}
\bibinfo{author}{Li, A.}, \bibinfo{author}{Li, Z.}, \bibinfo{author}{Bellotti,
  A.}, \bibinfo{year}{2023}.
\newblock \bibinfo{title}{Predicting loss given default of unsecured consumer
  loans with time-varying survival scores}.
\newblock \bibinfo{journal}{Pacific-Basin Finance Journal}
  \bibinfo{volume}{78}, \bibinfo{pages}{101949}.
\bibitem[{Olsen(1978)}]{olsen1978note}
\bibinfo{author}{Olsen, R.J.}, \bibinfo{year}{1978}.
\newblock \bibinfo{title}{Note on the uniqueness of the maximum likelihood
  estimator for the tobit model}.
\newblock \bibinfo{journal}{Econometrica: Journal of the Econometric Society} ,
  \bibinfo{pages}{1211--1215}.
\bibitem[{Ospina and Ferrari(2010)}]{ospina2010inflated}
\bibinfo{author}{Ospina, R.}, \bibinfo{author}{Ferrari, S.L.},
  \bibinfo{year}{2010}.
\newblock \bibinfo{title}{Inflated beta distributions}.
\newblock \bibinfo{journal}{Statistical papers} \bibinfo{volume}{51},
  \bibinfo{pages}{111--126}.
\bibitem[{Papke and Wooldridge(1996)}]{papke1996econometric}
\bibinfo{author}{Papke, L.E.}, \bibinfo{author}{Wooldridge, J.M.},
  \bibinfo{year}{1996}.
\newblock \bibinfo{title}{Econometric methods for fractional response variables
  with an application to 401 (k) plan participation rates}.
\newblock \bibinfo{journal}{Journal of applied econometrics}
  \bibinfo{volume}{11}, \bibinfo{pages}{619--632}.
\bibitem[{Qi and Yang(2009)}]{qi2009loss}
\bibinfo{author}{Qi, M.}, \bibinfo{author}{Yang, X.}, \bibinfo{year}{2009}.
\newblock \bibinfo{title}{Loss given default of high loan-to-value residential
  mortgages}.
\newblock \bibinfo{journal}{Journal of Banking \& Finance}
  \bibinfo{volume}{33}, \bibinfo{pages}{788--799}.
\bibitem[{Qi and Zhao(2011)}]{qi2011comparison}
\bibinfo{author}{Qi, M.}, \bibinfo{author}{Zhao, X.}, \bibinfo{year}{2011}.
\newblock \bibinfo{title}{Comparison of modeling methods for loss given
  default}.
\newblock \bibinfo{journal}{Journal of Banking \& Finance}
  \bibinfo{volume}{35}, \bibinfo{pages}{2842--2855}.
\bibitem[{Sigrist and Stahel(2011)}]{sigrist2011using}
\bibinfo{author}{Sigrist, F.}, \bibinfo{author}{Stahel, W.A.},
  \bibinfo{year}{2011}.
\newblock \bibinfo{title}{Using the censored gamma distribution for modeling
  fractional response variables with an application to loss given default}.
\newblock \bibinfo{journal}{ASTIN Bulletin: The Journal of the IAA}
  \bibinfo{volume}{41}, \bibinfo{pages}{673--710}.
\bibitem[{Tobin(1958)}]{tobin1958estimation}
\bibinfo{author}{Tobin, J.}, \bibinfo{year}{1958}.
\newblock \bibinfo{title}{Estimation of relationships for limited dependent
  variables}.
\newblock \bibinfo{journal}{Econometrica: journal of the Econometric Society} ,
  \bibinfo{pages}{24--36}.
\bibitem[{Yao et~al.(2017)Yao, Crook and Andreeva}]{yao2017enhancing}
\bibinfo{author}{Yao, X.}, \bibinfo{author}{Crook, J.},
  \bibinfo{author}{Andreeva, G.}, \bibinfo{year}{2017}.
\newblock \bibinfo{title}{Enhancing two-stage modelling methodology for loss
  given default with support vector machines}.
\newblock \bibinfo{journal}{European Journal of Operational Research}
  \bibinfo{volume}{263}, \bibinfo{pages}{679--689}.

\end{thebibliography}
\newpage
\appendix

\section{Tables }
\label{App:1}
\begin{table}[h]
\center
\caption{Correlation between CPI and Spot Rates and Transformations }\label{table:rates}
\scalebox{0.5}{
\begin{tabular}{|l|l|l|l|l|l|}
\hline
\textbf{Rates and Transformations} & \textbf{Correlation} & \textbf{Time} & \textbf{Category Description}                               & \textbf{Transformation} & \textbf{\begin{tabular}[c]{@{}l@{}}Stationary Test \\ (ADF) p- value\end{tabular}} \\ \hline
AVG\_FEDFNDTGT\_M.LDIFF12M         & 0.5215               & 2009-2022     &                                                             & LDIFF12M                & {\color[HTML]{009901} 0.1015}                                                      \\ \cline{1-3} \cline{5-6} 
AVG\_FEDFNDTGT\_M.RDIFF12M         & 0.5834               & 2009-2022     & \multirow{-2}{*}{Average Fed Funds Target}                  & RDIFF12M                & 0.8926                                                                             \\ \hline
COFI\_M                            & 0.5557               & 1978-2022     &                                                             & Raw                     & 0.7540                                                                             \\ \cline{1-3} \cline{5-6} 
COFI\_M.DIFF12M                    & 0.5088               & 1979-2022     & \multirow{-2}{*}{USD Cost of funding index (COFI)}          & DIFF12M                 & {\color[HTML]{009901} 0.0000}                                                      \\ \hline
FEDFND\_EFF\_M                     & 0.6861               & 1973-2022     & USD Fed Funds Effective - Month End                         & Raw                     & {\color[HTML]{009901} 0.1029}                                                      \\ \hline
MTA1Y\_M                           & 0.6491               & 1954-2022     & USD MTA 1-Year                                              & Raw                     & 0.3992                                                                             \\ \hline
PRIM\_MTG\_15Y\_M.LDIFF12M         & 0.5326               & 1992-2022     &                                                             & LDIFF12M                & {\color[HTML]{009901} 0.0067}                                                      \\ \cline{1-3} \cline{5-6} 
PRIM\_MTG\_15Y\_M.RDIFF12M         & 0.5628               & 1992-2022     & \multirow{-2}{*}{15-Year Primary Mortgage Rate}             & RDIFF12M                & {\color[HTML]{009901} 0.0129}                                                      \\ \hline
PRIM\_MTG\_30Y\_RT\_M              & 0.5878               & 1971-2022     &                                                             & Raw                     & 0.6022                                                                             \\ \cline{1-3} \cline{5-6} 
PRIM\_MTG\_30Y\_RT\_M.DIFF12M      & 0.5523               & 1972-2022     &                                                             & DIFF12M                 & {\color[HTML]{009901} 0.0005}                                                      \\ \cline{1-3} \cline{5-6} 
PRIM\_MTG\_30Y\_RT\_M.LDIFF12M     & 0.5146               & 1972-2022     & \multirow{-3}{*}{30-Year Primary Mortgage Rate}             & LDIFF12M                & {\color[HTML]{009901} 0.0004}                                                      \\ \hline
PRIM\_MTG\_51ARM\_M.RDIFF12M       & 0.5171               & 2006-2022     & 5/1 ARM Primary Mortgage Rate                               & RDIFF12M                & 0.4216                                                                             \\ \hline
PRIM\_MTG\_71ARM\_M.DIFF12M        & 0.5492               & 2006-2022     &                                                             & DIFF12M                 & 0.5249                                                                             \\ \cline{1-3} \cline{5-6} 
PRIM\_MTG\_71ARM\_M.LDIFF12M       & 0.5267               & 2006-2022     &                                                             & LDIFF12M                & 0.3284                                                                             \\ \cline{1-3} \cline{5-6} 
PRIM\_MTG\_71ARM\_M.RDIFF12M       & 0.5410               & 2006-2022     & \multirow{-3}{*}{7/1 ARM Primary Mortgage Rate}             & RDIFF12M                & 0.6758                                                                             \\ \hline
PRIM\_MTG\_JUMBFX15\_M.DIFF12M     & 0.5838               & 2006-2022     &                                                             & DIFF12M                 & 0.7159                                                                             \\ \cline{1-3} \cline{5-6} 
PRIM\_MTG\_JUMBFX15\_M.LDIFF12M    & 0.5853               & 2006-2022     &                                                             & LDIFF12M                & 0.5817                                                                             \\ \cline{1-3} \cline{5-6} 
PRIM\_MTG\_JUMBFX15\_M.RDIFF12M    & 0.6079               & 2006-2022     & \multirow{-3}{*}{15-Year Jumbo Fixed Primary Mortgage Rate} & RDIFF12M                & 0.8151                                                                             \\ \hline
PRIM\_MTG\_JUMBFX30\_M.DIFF12M     & 0.6422               & 2006-2022     &                                                             & DIFF12M                 & 0.4404                                                                             \\ \cline{1-3} \cline{5-6} 
PRIM\_MTG\_JUMBFX30\_M.LDIFF12M    & 0.6459               & 2006-2022     &                                                             & LDIFF12M                & 0.3076                                                                             \\ \cline{1-3} \cline{5-6} 
PRIM\_MTG\_JUMBFX30\_M.RDIFF12M    & 0.6511               & 2006-2022     & \multirow{-3}{*}{30-Year Jumbo Fixed Primary Mortgage Rate} & RDIFF12M                & 0.4822                                                                             \\ \hline
SCND\_15Y\_MTG\_M.LDIFF12M         & 0.6015               & 1993-2022     &                                                             & LDIFF12M                & {\color[HTML]{009901} 0.0001}                                                      \\ \cline{1-3} \cline{5-6} 
SCND\_15Y\_MTG\_M.RDIFF12M         & 0.6485               & 1993-2022     & \multirow{-2}{*}{15-Year Secondary Mortgage Rate}           & RDIFF12M                & {\color[HTML]{009901} 0.0001}                                                      \\ \hline
SCND\_MTG\_30Y\_M.DIFF12M          & 0.5115               & 1993-2022     &                                                             & DIFF12M                 & {\color[HTML]{009901} 0.0066}                                                      \\ \cline{1-3} \cline{5-6} 
SCND\_MTG\_30Y\_M.LDIFF12M         & 0.6185               & 1993-2022     &                                                             & LDIFF12M                & {\color[HTML]{009901} 0.0001}                                                      \\ \cline{1-3} \cline{5-6} 
SCND\_MTG\_30Y\_M.RDIFF12M         & 0.6576               & 1993-2022     & \multirow{-3}{*}{30-Year Secondary Mortgage Rate}           & RDIFF12M                & {\color[HTML]{009901} 0.0005}                                                      \\ \hline
SWP10Y\_M.LDIFF12M                 & 0.5169               & 1992-2022     & USD 10 Year Swap Rate                                       & LDIFF12M                & {\color[HTML]{009901} 0.0003}                                                      \\ \hline
SWP15Y\_M.LDIFF112M                & 0.5004               & 1992-2022     &                                                             & LDIFF12M                & {\color[HTML]{009901} 0.0006}                                                      \\ \cline{1-3} \cline{5-6} 
SWP15Y\_M.RDIFF12M                 & 0.5326               & 1992-2022     & \multirow{-2}{*}{USD 15 Year Swap Rate}                     & RDIFF12M                & {\color[HTML]{009901} 0.0234}                                                      \\ \hline
SWP20Y\_M.RDIFF12M                 & 0.5136               & 1992-2022     & USD 20 Year Swap Rate                                       & RDIFF12M                & {\color[HTML]{009901} 0.0184}                                                      \\ \hline
SWP2Y\_M.LDIFF12M                  & 0.5848               & 1992-2022     &                                                             & LDIFF12M                & {\color[HTML]{009901} 0.0000}                                                      \\ \cline{1-3} \cline{5-6} 
SWP2Y\_M.RDIFF12M                  & 0.6734               & 1992-2022     & \multirow{-2}{*}{USD 2-Year Swap Rate}                      & RDIFF12M                & {\color[HTML]{009901} 0.0004}                                                      \\ \hline
SWP3Y\_M.LDIFF12M                  & 0.5768               & 1992-2022     &                                                             & LDIFF12M                & {\color[HTML]{009901} 0.0000}                                                      \\ \cline{1-3} \cline{5-6} 
SWP3Y\_M.RDIFF12M                  & 0.7031               & 1992-2022     & \multirow{-2}{*}{USD 3-Year Swap Rate}                      & RDIFF12M                & {\color[HTML]{009901} 0.0000}                                                      \\ \hline
SWP4Y\_M.LDIFF12M                  & 0.5603               & 1992-2022     &                                                             & LDIFF12M                & {\color[HTML]{009901} 0.0000}                                                      \\ \cline{1-3} \cline{5-6} 
SWP4Y\_M.RDIFF12M                  & 0.6822               & 1992-2022     & \multirow{-2}{*}{USD 4-Year Swap Rate}                      & RDIFF12M                & {\color[HTML]{009901} 0.0000}                                                      \\ \hline
SWP5Y\_M.LDIFF12M                  & 0.5481               & 1992-2022     &                                                             & LDIFF12M                & {\color[HTML]{009901} 0.0000}                                                      \\ \cline{1-3} \cline{5-6} 
SWP5Y\_M.RDIFF12M                  & 0.6500               & 1992-2022     & \multirow{-2}{*}{USD 5-Year Swap Rate}                      & RDIFF12M                & {\color[HTML]{009901} 0.0002}                                                      \\ \hline
SWP6Y\_M.LDIFF12M                  & 0.5397               & 1992-2022     &                                                             & LDIFF12M                & {\color[HTML]{009901} 0.0001}                                                      \\ \cline{1-3} \cline{5-6} 
SWP6Y\_M.RDIFF12M                  & 0.6237               & 1992-2022     & \multirow{-2}{*}{USD 6-Year Swap Rate}                      & RDIFF12M                & {\color[HTML]{009901} 0.0019}                                                      \\ \hline
SWP7Y\_M.LDIFF12M                  & 0.5327               & 1992-2022     &                                                             & LDIFF12M                & {\color[HTML]{009901} 0.0001}                                                      \\ \cline{1-3} \cline{5-6} 
SWP7Y\_M.RDIFF12M                  & 0.6028               & 1992-2022     & \multirow{-2}{*}{USD 7-Year Swap Rate}                      & RDIFF12M                & {\color[HTML]{009901} 0.0047}                                                      \\ \hline
TSY10Y\_M.RDIFF12M                 & 0.5639               & 1992-2022     & USD 10-Year Swap Rate                                       & RDIFF12M                & 0.0180                                                                             \\ \hline
TSY1Y\_M.DIFF12M                   & 0.6370               & 2009-2022     &                                                             & DIFF12M                 & {\color[HTML]{009901} 0.0813}                                                      \\ \cline{1-3} \cline{5-6} 
TSY1Y\_M.LDIF12M                   & 0.5732               & 2009-2022     &                                                             & LDIFF12M                & {\color[HTML]{009901} 0.0462}                                                      \\ \cline{1-3} \cline{5-6} 
TSY1Y\_M.RDIFF12M                  & 0.7034               & 2009-2022     & \multirow{-3}{*}{USD 1-Year Treasury Rate}                  & RDIFF12M                & 0.5267                                                                             \\ \hline
TSY20Y\_M.RDIFF12M                 & 0.5223               & 1992-2022     & USD 20-Year Treasury Rate                                   & RDIFF12M                & 0.1803                                                                             \\ \hline
TSY2Y\_M.RDIFF12M                  & 0.5985               & 1990-2022     & USD 2-Year Treasury Rate                                    & RDIFF12M                & {\color[HTML]{009901} 0.0025}                                                      \\ \hline
TSY3Y\_M.RDIFF12M                  & 0.5941               & 1989-2022     & USD 3-Year Treasury Rate                                    & RDIFF12M                & {\color[HTML]{009901} 0.0000}                                                      \\ \hline
TSY5Y\_M.RDIFF12M                  & 0.5298               & 1986-2022     & USD 5-Year Treasury Rate                                    & RDIFF12M                & {\color[HTML]{009901} 0.0001}                                                      \\ \hline
TSY6M\_M.RDIFF12M                  & 0.5821               & 1996-2022     & USD 6-Month Treasury Rate                                   & RDIFF12M                & {\color[HTML]{009901} 0.0000}                                                      \\ \hline
UER\_M               & 0.0910             & 1950-2022     & National Unemployment Rate                                 & Raw               & {\color[HTML]{009901} 0.0035}                                                      \\ \hline
\end{tabular}
}
\end{table}

\begin{table}[h]
\center
\caption{Correlation between CPI and 1Y Forward Rates and/or Rates Transformations}\label{table:forward_rates}
\scalebox{0.5}{
\begin{tabular}{|l|c|c|l|l|c|}
\hline
\textbf{\begin{tabular}[c]{@{}l@{}}Rates and Transformations \\ 1 Yr Forward\end{tabular}} & \textbf{Correlation} & \textbf{Time} & \textbf{Y Category Descriptions}          & \textbf{Transformations} & \textbf{\begin{tabular}[c]{@{}c@{}}Stationary Test \\ (ADF) P-value\end{tabular}} \\ \hline
\cellcolor[HTML]{FFCCC9}CPI\_M.RDIFF12M                                                    & 0.7254               & 1952-2022     & CPI                                       & RDIFF12M                 & {\color[HTML]{32CB00} 0.0147}                                                     \\ \hline
\cellcolor[HTML]{FFCCC9}PRIM\_MTG\_30Y\_RT\_M                                              & 0.7116               & 1971-2022     & 30-Year Primary Mortgage Rate             & Raw                      & 0.6022                                                                            \\ \hline
\cellcolor[HTML]{FFCCC9}COFI\_M                                                            & 0.7064               & 1978-2022     & USD Cost of funding index (COFI)          & Raw                      & 0.7540                                                                            \\ \hline
\cellcolor[HTML]{FFFFC7}FEDFND\_EFF\_M                                                     & 0.696                & 1973-2022     & USD Fed Funds Effective Month End         & Raw                      & {\color[HTML]{32CB00} 0.1029}                                                     \\ \hline
\cellcolor[HTML]{FFCCC9}AVG\_FEDFNDTGT\_M.RDIFF12M                                         & 0.6244               & 2009-2022     & Average Fed Funds Target                  & RDIFF12M                 & 0.8926                                                                            \\ \hline
\cellcolor[HTML]{FFCCC9}TSY10Y\_M                                                          & 0.6205               & 1982-2022     & USD 10-Year Treasury Rate                 & Raw                      & {\color[HTML]{32CB00} 0.0233}                                                     \\ \hline
\cellcolor[HTML]{FFCCC9}TSY30Y\_M                                                          & 0.6091               & 1982-2022     & USD 30-Year Treasury Rate                 & Raw                      & {\color[HTML]{32CB00} 0.0323}                                                     \\ \hline
\cellcolor[HTML]{FFCCC9}TSY3Y\_M                                                           & 0.5472               & 1988-2022     & USD 3-Year Treasury Rate                  & Raw                      & 0.3411                                                                            \\ \hline
\cellcolor[HTML]{FFCCC9}TSY2Y\_M                                                           & 0.5285               & 1989-2022     & USD 2-Year Treasury Rate                  & Raw                      & 0.1424                                                                            \\ \hline
\cellcolor[HTML]{FFCCC9}PRIM\_MTG\_51ARM\_M                                                & 0.5241               & 2005-2022     & 5/1 ARM Primary Mortgage Rate             & Raw                      & 0.6276                                                                            \\ \hline
\cellcolor[HTML]{FFCCC9}PRIM\_MTG\_71ARM\_M                                                & 0.5113               & 2005-2022     & 7/1 ARM Primary Mortgage Rate             & Raw                      & 0.5749                                                                            \\ \hline
\cellcolor[HTML]{FFCCC9}TSY5Y\_M                                                           & 0.5065               & 1985-2022     & USD 5-Year Treasury Rate                  & Raw                      & 0.4789                                                                            \\ \hline
\cellcolor[HTML]{FFFFC7}TSY1Y\_M.RDIFF12M                                                  & 0.5031               & 2009-2022     & USD 1-Year Treasury Rate                  & RDIFF12M                 & 0.5267                                                                            \\ \hline
\cellcolor[HTML]{FFCCC9}TSY7Y\_M                                                           & 0.4946               & 1984-2022     & USD 7-Year Treasury Rate                  & Raw                      & {\color[HTML]{32CB00} 0.0081}                                                     \\ \hline
\cellcolor[HTML]{FFFFC7}MEDN\_INCM\_M.RDIFF12M                                             & 0.4895               & 1968-2022     & National Median Income                    & RDIFF12M                 & 0.1824                                                                            \\ \hline
\cellcolor[HTML]{FFFFC7}PRIM\_MTG\_51ARM\_M.RDIFF12M                                       & 0.4847               & 2006-2022     & 5/1 ARM Primary Mortgage Rate             & RDIFF12M                 & 0.4216                                                                            \\ \hline
\cellcolor[HTML]{FFFFC7}NMNL\_DISP\_INCM\_M.RDIFF12M                                       & 0.4774               & 1960-2022     & US nominal disposable income levels       & RDIFF12M                 & 0.2772                                                                            \\ \hline
\cellcolor[HTML]{FFFFC7}PRIM\_MTG\_71ARM\_M.RDIFF12M                                       & 0.4739               & 2006-2022     & 7/1 ARM Primary Mortgage Rate             & RDIFF12M                 & 0.6758                                                                            \\ \hline
\cellcolor[HTML]{FFFFC7}BBB\_YLD\_M.RDIFF12M                                               & 0.4661               & 2000-2022     & BBB Corporate Yield                       & RDIFF12M                 & {\color[HTML]{32CB00} 0.0772}                                                     \\ \hline
\cellcolor[HTML]{FFCCC9}PRIM\_MTG\_15Y\_M                                                  & 0.464                & 1991-2022     & 15-Year Primary Mortgage Rate             & Raw                      & 0.3222                                                                            \\ \hline
\cellcolor[HTML]{FFCCC9}TSY1Y\_M                                                           & 0.4588               & 2008-2022     & USD 1-Year Treasury Rate                  & Raw                      & 0.5921                                                                            \\ \hline
\cellcolor[HTML]{FFCCC9}PRIM\_MTG\_JUMBFX15\_M                                             & 0.4493               & 2005-2022     & 15-Year Jumbo Fixed Primary Mortgage Rate & Raw                      & 0.6519                                                                            \\ \hline
\cellcolor[HTML]{FFCCC9}TSY20Y\_M                                                          & 0.4476               & 1991-2022     & USD 20-Year Treasury Rate                 & Raw                      & 0.2712                                                                            \\ \hline
\cellcolor[HTML]{FFFFC7}PRIM\_MTG\_JUMBFX30\_M.RDIFF12M                                    & 0.4455               & 2006-2022     & 30-Year Jumbo Fixed Primary Mortgage Rate & RDIFF12M                 & 0.4822                                                                            \\ \hline
\cellcolor[HTML]{FFCCC9}PRIM\_MTG\_JUMBFX30\_M                                             & 0.4439               & 2005-2022     & 30-Year Jumbo Fixed Primary Mortgage Rate & Raw                      & 0.6184                                                                            \\ \hline
\cellcolor[HTML]{FFFFC7}PRIM\_MTG\_JUMBFX15\_M.RDIFF12M                                    & 0.4422               & 2006-2022     & 15-Year Jumbo Fixed Primary Mortgage Rate & RDIFF12M                 & 0.8151                                                                            \\ \hline
\cellcolor[HTML]{FFCCC9}SWP4Y\_M                                                           & 0.4421               & 1991-2022     & USD 4-Year Swap Rate                      & Raw                      & 0.1865                                                                            \\ \hline
\cellcolor[HTML]{FFCCC9}SWP5Y\_M                                                           & 0.441                & 1991-2022     & USD 5-Year Swap Rate                      & Raw                      & 0.1744                                                                            \\ \hline
\cellcolor[HTML]{FFCCC9}SWP3Y\_M                                                           & 0.4404               & 1991-2022     & USD 3-Year Swap Rate                      & Raw                      & 0.1980                                                                            \\ \hline
\cellcolor[HTML]{FFCCC9}SWP6Y\_M                                                           & 0.4392               & 1991-2022     & USD 6-Year Swap Rate                      & Raw                      & 0.2749                                                                            \\ \hline
\cellcolor[HTML]{DAE8FC}PRIM\_MTG\_51ARM\_M.DIFF12M                                        & 0.4379               & 2006-2022     & 5/1 ARM Primary Mortgage Rate             & DIFF12M                  & 0.5127                                                                            \\ \hline
\cellcolor[HTML]{FFCCC9}SWP7Y\_M                                                           & 0.4369               & 1991-2022     & USD 7-Year Swap Rate                      & Raw                      & 0.2698                                                                            \\ \hline
\cellcolor[HTML]{DAE8FC}CDX\_HY\_5Y\_SP\_M.DIFF12M                                         & 0.4356               & 2002-2022     & CDX High Yield 5-Year Spread              & DIFF12M                  & {\color[HTML]{32CB00} 0.0020}                                                     \\ \hline
\cellcolor[HTML]{FFCCC9}SWP2Y\_M                                                           & 0.4347               & 1991-2022     & USD 2-Year Swap Rate                      & Raw                      & 0.2421                                                                            \\ \hline
\cellcolor[HTML]{FFCCC9}SWP10Y\_M                                                          & 0.4292               & 1991-2022     & USD 10-Year Swap Rate                     & Raw                      & 0.2840                                                                            \\ \hline
\cellcolor[HTML]{DAE8FC}BBB\_YLD\_M.DIFF12M                                                & 0.4146               & 2000-2022     & BBB Corporate Yield                       & DIFF12M                  & {\color[HTML]{32CB00} 0.0279}                                                     \\ \hline
\cellcolor[HTML]{DAE8FC}PRIM\_MTG\_71ARM\_M.DIFF12M                                        & 0.4139               & 2006-2022     & 7/1 ARM Primary Mortgage Rate             & DIFF12M                  & 0.5249                                                                            \\ \hline
\cellcolor[HTML]{FFCCC9}SWP15Y\_M                                                          & 0.4138               & 1991-2022     & USD 15-Year Swap Rate                     & Raw                      & 0.3181                                                                            \\ \hline
\cellcolor[HTML]{FFCCC9}SCND\_15Y\_MTG\_M                                                  & 0.4105               & 1992-2022     & 15-Year Secondary Mortgage Rate           & Raw                      & 0.3585                                                                            \\ \hline
\cellcolor[HTML]{DAE8FC}AVG\_FEDFNDTGT\_M.DIFF12M                                          & 0.4038               & 2009-2022     & Average Fed Funds Target                  & DIFF12M                  & 0.6136                                                                            \\ \hline
\cellcolor[HTML]{FFFFC7}TSY6M\_M.RDIFF12M                                                  & 0.385                & 1996-2022     & USD 6-Month Treasury Rate                 & RDIFF12M                 & {\color[HTML]{32CB00} 0.0000}                                                     \\ \hline
\cellcolor[HTML]{FFCCC9}AVG\_FEDFNDTGT\_M                                                  & 0.3818               & 2008-2022     & Average Fed Funds Target                  & Raw                      & 0.4359                                                                            \\ \hline
\cellcolor[HTML]{DAE8FC}PRIM\_MTG\_JUMBFX15\_M.DIFF12M                                     & 0.3811               & 2006-2022     & 15-Year Jumbo Fixed Primary Mortgage Rate & DIFF12M                  & 0.7159                                                                            \\ \hline
\cellcolor[HTML]{FFFFC7}CDX\_HY\_5Y\_SP\_M.RDIFF12M                                        & 0.3804               & 2002-2022     & CDX High Yield 5-Year Spread              & RDIFF12M                 & {\color[HTML]{32CB00} 0.0135}                                                     \\ \hline
PRIM\_MTG\_JUMBFX30M.DIFF12N                                                               & 0.3722               & 2006-2022     & 30-Year Jumbo Fixed Primary Mortgage Rate & DIFF12M                  & 0.4404                                                                            \\ \hline
\cellcolor[HTML]{FFFFC7}BBB\_SP\_M.RDIFF12M                                                & 0.3527               & 2000-2022     & BBB Corporate Yield                       & RDIFF12M                 & {\color[HTML]{32CB00} 0.0025}                                                     \\ \hline
\cellcolor[HTML]{DAE8FC}TSY1Y\_M.DIFF12M                                                   & 0.3316               & 2009-2022     & USD 1-Year Treasury Rate                  & DIFF12M                  & {\color[HTML]{32CB00} 0.0813}                                                     \\ \hline  
\cellcolor[HTML]{FFCCC9}UER\_M                      & 0.3292              & 1950-2022     & National Unemployment Rate             & Raw            &   {\color[HTML]{32CB00} 0.0035}       \\ \hline
\cellcolor[HTML]{FFFFC7}TSY1M\_M.RDIFF12M                                                  & 0.3194               & 2008-2022     & USD 1-Month Treasury Rate                 & RDIFF12M                 & 0.8452                                                                            \\ \hline
\cellcolor[HTML]{FFFFC7}PRIM\_MTG\_15Y\_M.RDIFF12M                                         & 0.3191               & 1992-2022     & 15-Year Primary Mortgage Rate             & RDIFF12M                 & {\color[HTML]{32CB00} 0.0129}                                                     \\ \hline
\cellcolor[HTML]{DAE8FC}CDX\_IG\_5Y\_SP\_M.DIFF12M                                         & 0.3166               & 2004-2022     & CDX Investment Grade 5-Year Spreads       & DIFF12M                  & {\color[HTML]{32CB00} 0.0191}                                                     \\ \hline
\cellcolor[HTML]{FFFFC7}TSY3M\_M.RDIFF12M                                                  & 0.307                & 1996-2022     & USD 3-Month Treasury Rate                 & RDIFF12M                 & 0.8554                                                                            \\ \hline
\cellcolor[HTML]{DAE8FC}COFI\_M.DIFF12M                                                    & 0.3045               & 1979-2022     & USD Cost of funding index (COFI)          & DIFF12M                  & {\color[HTML]{32CB00} 0.0000}                                                     \\ \hline
\cellcolor[HTML]{FFFFC7}SCND\_15Y\_MTG\_M.RDIFF12M                                         & 0.3035               & 1993-2022     & 15-Year Secondary Mortgage Rate           & RDIFF12M                 & {\color[HTML]{32CB00} 0.0001}                                                     \\ \hline
\cellcolor[HTML]{FFFFC7}SWP2Y\_M.RDIFF12M                                                  & 0.2651               & 1992-2022     & USD 2-Year Swap Rate                      & RDIFF12M                 & {\color[HTML]{32CB00} 0.0004}                                                     \\ \hline
\cellcolor[HTML]{DAE8FC}PRIM\_MTG\_30Y\_RT\_M.DIFF12M                                      & 0.2548               & 1972-2022     & 30-Year Primary Mortgage Rate             & DIFF12M                  & {\color[HTML]{32CB00} 0.0005}                                                     \\ \hline
\cellcolor[HTML]{DAE8FC}UNEMP\_M.DIFF12M                                                   & 0.2498               & 1951-2022     & US Unemployment Rate                      & DIFF12M                  & {\color[HTML]{32CB00} 0.0000}                                                     \\ \hline
\cellcolor[HTML]{DAE8FC}AA\_SP\_M.DIFF12M                                                  & 0.2259               & 2000-2022     & AA Corporate Spreads                      & DIFF12M                  & {\color[HTML]{32CB00} 0.0016}                                                     \\ \hline
\cellcolor[HTML]{DAE8FC}MTG\_SP\_30Y\_M.DIFF12M                                            & 0.2196               & 1993-2022     & 30-Year Primary/Secondary Mortgage Spread & DIFF12M                  & {\color[HTML]{32CB00} 0.0002}                                                     \\ \hline
\cellcolor[HTML]{DAE8FC}MTG\_SP\_15Y\_M.DIFF12M                                            & 0.1592               & 1993-2022     & 15-Year Primary/Secondary Mortgage Spread & DIFF12M                  & {\color[HTML]{32CB00} 0.0002}                                                     \\ \hline
\cellcolor[HTML]{DAE8FC}PRIM\_MTG\_15Y\_M.DIFF12M                                          & 0.1484               & 1992-2022     & 15-Year Primary Mortgage Rate             & DIFF12M                  & {\color[HTML]{32CB00} 0.0180}                                                     \\ \hline
\cellcolor[HTML]{FFFFC7}SP500\_M.RDIFF12M                                                  & -0.1077              & 1951-2022     & S\&P500 Index                             & RDIFF12M                 & {\color[HTML]{32CB00} 0.0000}                                                     \\ \hline
\cellcolor[HTML]{FFFFC7}DJIA\_TOT\_M.RDIFF12M                                              & -0.2615              & 1988-2022     & Dow Jones Total Stock Market Index (EOP)  & RDIFF12M                 & {\color[HTML]{32CB00} 0.0028}                                                     \\ \hline
\end{tabular}
}
\end{table} 

\begin{table}[h]
\centering
\caption{Correlation between CPI YoY and CoStar Market Sales Price Indices YoY 1 Year Later, monthly data }
\label{table: Costar_CPI}
\scalebox{0.8}{
\begin{tabular}{|c|c|l|r|}
\hline
\multicolumn{1}{|c|}{\textbf{Time}}                                                   & \multicolumn{1}{c|}{\textbf{MSA}} & \textbf{Prop} & \textbf{Correlation}          \\ \hline\hline
                                                                                                        &                                   & APT           & -0.3505                       \\ \cline{3-4} 
                                                                                                        & \multirow{-2}{*}{US (National)}   & OFF           & -0.3346                       \\ \cline{2-4} 
                                                                                                        &                                   & APT           & -0.3216                       \\ \cline{3-4} 
                                                                                                        & \multirow{-2}{*}{New York}        & OFF           & -0.2699                       \\ \cline{2-4} 
                                                                                                        &                                   & APT           & -0.2692                       \\ \cline{3-4} 
\multirow{-6}{*}{\begin{tabular}[c]{@{}c@{}}All Years \\ (1983-03 to 2022-12)\end{tabular}}             & \multirow{-2}{*}{LA}              & OFF           & -0.3098                 \\ \hline\hline
                                                                                                        &                                   & APT           & -0.4125                       \\ \cline{3-4} 
                                                                                                        & \multirow{-2}{*}{US (National)}   & OFF           & -0.2230                       \\ \cline{2-4} 
                                                                                                        &                                   & APT           & -0.3960                       \\ \cline{3-4} 
                                                                                                        & \multirow{-2}{*}{New York}        & OFF           & -0.1354                       \\ \cline{2-4} 
                                                                                                        &                                   & APT           & -0.3776                       \\ \cline{3-4} 
\multirow{-6}{*}{\begin{tabular}[c]{@{}c@{}}Modeling Period \\ (2003-12 to 2019-12)\end{tabular}}       & \multirow{-2}{*}{LA}              & OFF           & -0.2199             \\ \hline\hline
                                                                                                        &                                   & APT           & -0.8579                       \\ \cline{3-4} 
                                                                                                        & \multirow{-2}{*}{US (National)}   & OFF           & -0.6284                       \\ \cline{2-4} 
                                                                                                        &                                   & APT           & -0.8271                       \\ \cline{3-4} 
                                                                                                        & \multirow{-2}{*}{New York}        & OFF           & -0.5106                       \\ \cline{2-4} 
                                                                                                        &                                   & APT           & -0.8577                       \\ \cline{3-4} 
\multirow{-6}{*}{\begin{tabular}[c]{@{}c@{}}GFC \\ (2008-04 to 2010-03)\end{tabular}}                   & \multirow{-2}{*}{LA}              & OFF           & -0.6029                  \\ \hline\hline
                                                                                                        &                                   & APT           & -0.5776                       \\ \cline{3-4} 
                                                                                                        & \multirow{-2}{*}{US (National)}   & OFF           & -0.5483                       \\ \cline{2-4} 
                                                                                                        &                                   & APT           & -0.5668                       \\ \cline{3-4} 
                                                                                                        & \multirow{-2}{*}{New York}        & OFF           & -0.6873                       \\ \cline{2-4} 
                                                                                                        &                                   & APT           & -0.5762                       \\ \cline{3-4} 
\multirow{-6}{*}{\begin{tabular}[c]{@{}c@{}}Early 1990s Recession \\ (1990-07 to 1991-03)\end{tabular}} & \multirow{-2}{*}{LA}              & OFF           & -0.5804            \\ \hline\hline
                                                                                                        &                                   & APT           & -0.5889                       \\ \cline{3-4} 
                                                                                                        & \multirow{-2}{*}{US (National)}   & OFF           & {\color[HTML]{FE0000} 0.3125} \\ \cline{2-4} 
                                                                                                        &                                   & APT           & -0.5977                       \\ \cline{3-4} 
                                                                                                        & \multirow{-2}{*}{New York}        & OFF           & {\color[HTML]{FE0000} 0.5569} \\ \cline{2-4} 
                                                                                                        &                                   & APT           & -0.5730                       \\ \cline{3-4} 
\multirow{-6}{*}{\begin{tabular}[c]{@{}c@{}}COVID-19 Period \\ (2020-01 to 2021-12)\end{tabular}}       & \multirow{-2}{*}{LA}              & OFF           & {\color[HTML]{FE0000} 0.2579} \\ \hline
\end{tabular}
}
\end{table}

\begin{table}[ht]
\centering
\caption{Correlation between CPI YoY and CoStar Market Sales Price Index YoY 1 Year Later, monthly data, by CPI YoY Buckets }
\label{table: Costar_CPI_byBucket}
\scalebox{0.75}{
\begin{tabular}{|c|c|l|l|r|}
\hline
\multicolumn{1}{|l|}{\textbf{CPI YoY Bucket}} & \textbf{Month Count}       & \textbf{MSA}                   & \textbf{Prop} & \multicolumn{1}{l|}{\textbf{Correlation}} \\ \hline
\multirow{6}{*}{\textless{}=0.02}             & \multirow{6}{*}{150} & \multirow{2}{*}{US (National)} & APT           & -0.1901                                   \\ \cline{4-5} 
                                              &                      &                                & OFF           & 0.2091                                    \\ \cline{3-5} 
                                              &                      & \multirow{2}{*}{New York}      & APT           & -0.1710                                   \\ \cline{4-5} 
                                              &                      &                                & OFF           & 0.3253                                    \\ \cline{3-5} 
                                              &                      & \multirow{2}{*}{LA}            & APT           & -0.0782                                   \\ \cline{4-5} 
                                              &                      &                                & OFF           & 0.1615                                    \\ \hline\hline
\multirow{6}{*}{0.02 - 0.04}                  & \multirow{6}{*}{243} & \multirow{2}{*}{US (National)} & APT           & -0.2670                                   \\ \cline{4-5} 
                                              &                      &                                & OFF           & -0.3116                                   \\ \cline{3-5} 
                                              &                      & \multirow{2}{*}{New York}      & APT           & -0.1766                                   \\ \cline{4-5} 
                                              &                      &                                & OFF           & -0.1398                                   \\ \cline{3-5} 
                                              &                      & \multirow{2}{*}{LA}            & APT           & -0.1586                                   \\ \cline{4-5} 
                                              &                      &                                & OFF           & -0.1755                                   \\ \hline\hline
\multirow{6}{*}{\textgreater 0.04}            & \multirow{6}{*}{85}  & \multirow{2}{*}{US (National)} & APT           & 0.1931                                    \\ \cline{4-5} 
                                              &                      &                                & OFF           & 0.0221                                    \\ \cline{3-5} 
                                              &                      & \multirow{2}{*}{New York}      & APT           & 0.0232                                    \\ \cline{4-5} 
                                              &                      &                                & OFF           & -0.1847                                   \\ \cline{3-5} 
                                              &                      & \multirow{2}{*}{LA}            & APT           & 0.1318                                    \\ \cline{4-5} 
                                              &                      &                                & OFF           & -0.0995                                   \\ \hline
\end{tabular}
}

\end{table}

\newpage




\end{document}